\documentclass[aps,pra,showpacs,superscriptaddress,twoside,twocolumn,10pt]{revtex4-1}
\usepackage[colorlinks=true, citecolor=cyan, urlcolor=teal ]{hyperref}
\usepackage{times,epsfig,amssymb,amsfonts,amsmath,bm,subfigure,mathtools,amsthm,
braket,soul,enumitem,xcolor,physics,graphics,graphicx,array,multirow}

\newtheorem{definition}{Definition}

\begin{document}
\title{Noisy dynamics of Gaussian entanglement: a transient
bound entangled phase before separability}
\author{Gurvir Singh}
\email{gurvirsingh@iisermohali.ac.in}
\affiliation{Department of Physical Sciences, Indian
Institute of Science Education and Research Mohali, Sector
81 SAS Nagar, Punjab 140306, India}  
\author{Saptarshi Roy}
\email{sapsoy@gmail.com}
\affiliation{The University of Hong Kong, Pokfulam Road, Hong Kong}  
\author{Arvind}  
\email{arvind@iisermohali.ac.in}
\affiliation{Department of Physical Sciences, Indian
Institute of Science Education and Research Mohali, Sector
81 SAS Nagar, Punjab 140306, India}  
\begin{abstract}
We discover a new class of Gaussian bound entangled
states of four-mode continuous-variable systems. These
states appear as a transient phase when certain
NPT-entangled Gaussian states are evolved under a noisy
environment. A thermal bath comprising of harmonic
oscillators is allowed to interact with one or modes of the
system and a wide variety of initial Gaussian entangled (NPT
as well as PPT) states are studied.  The robustness of
entanglement is defined as the time duration for which the
entanglement of the initial state is preserved under the
noisy dynamics.  We access the separability by utilizing
standard semi-definite programming techniques.  While most
states lose their entanglement after a certain time across
all bipartitions, an exception is observed for a
three-parameter family of states which we call the
generalized four-mode squeezed vacuum (gFMSV) states, 
which transitions to a bound entangled state, and remains so
for a finite window of time. This dynamical onset of bound
entanglement in continuous-variable systems is the central
observation of our work. We carry out the analysis for 
Haar-random four-mode states (both pure and mixed) to  scan the
state space for transient bound entangled phase.
\end{abstract}
\maketitle
\section{Introduction}
The field of continuous variable (CV) quantum information has
significantly evolved over time,
theoretically as well as
experimentally~\cite{Weedbrook_2012,Adesso_2014}. Gaussian states of CV
systems have gathered maxium attention due to their elegant
mathematical description and
feasibility~\cite{Adesso_2014}. From the $1990$s
applications of phase space methods brought out a rich
structure that lead to important developments in the
area~\cite{Arvind_1995,arvind_two_mode}.  Notably, the PPT criterion was
reformulated for CV-systems~\cite{PhysRevLett.84.2722} where
the partial transposition could be interpreted as partial
time reversal~\cite{PhysRevLett.84.2726}.
Bell inequalities have also been formulated
and studied for CV
systems~\cite{arvind_bell_cv1,arvind_bell_cv2}.  The focus
of
several other works in the CV sector was to cast protocols
that were devised in the discrete case into the CV
framework.  Some notable works range from communication
protocols like dense coding~\cite{cvdc1,cvdc2} and
teleportation~\cite{cvtel1,Pirandola2006,cvtel3}, to
analyzing quantum entanglement~\cite{Lami_2018} and other
quantum correlations~\cite{gqd}.
Proposals for developing quantum technologies using CV
quantum systems include quantum computation and machine
learning~\cite{Lau2017,weed,seth2019}, teleportation~\cite{sharma2025} to quantum key
distribution (CV-QKD) over long distances
~\cite{Laudenbach2018,Pirandola2024,Zhang_2024}.
Non-trivial CV
states with a quantum advantage can be generated using
the nonlinear interaction of a crystal with a
laser~\cite{exp}. While there are a number of similarities
between the DV and CV systems, there are also inherent
differences owing to the infinite dimensionality of CV
systems and the subtle aspects of quantum light that they are
realized through.

The entanglement structure and existence of bound
entanglement which concern us in this paper are quite
different in the CV systems as opposed to the DV
systems~\cite{Horodecki_2000}.  While in DV systems bound
entanglement is ubiquitous in $3\otimes 3$ and higher-dimensional
bipartite and multipartite systems, for CV systems within the family of
Gaussian states, bound entanglement is a rare
phenomenon~\cite{horodecki2001boundentanglementcontinuousvariables}.
As was pointed out by Werner and
Wolf~\cite{PhysRevLett.86.3658}, the simplest Gaussian case
where bound entangled states can exist is in the case of
$2+2$ modes.  While some useful applications have been found
for DV-bound entangled states such as superactivation of
bound entanglement which involves using tensor products of
bound entangled states to generate distillable states
\cite{PhysRevLett.90.107901}, positive key
rate~\cite{PhysRevA.75.032306,PhysRevA.102.032415} and
quantum metrology~\cite{PhysRevLett.120.020506}, the
usefulness of Gaussian bound entanglement is yet to be
explored~\cite{Navascu_s_2005,PhysRevA.72.012303}. This
motivates the necessity to carry out further investigations
on Gaussian bound entanglement.

It is important to study the effect of environmental
interactions on states, since noise is ubiquitous. In this
work, we investigate the dynamics of both distillable and
bound entangled Gaussian states of CV systems under the
influence of environmental interactions. In the case of CV
systems, most earlier investigations revolve around the
evolution of two-mode Gaussian states in different kinds of
dissipative environments and thus bound entanglement remain
out of the scope of these
investigations~\cite{refId0,marian2015decay,xiang2011,PhysRevA.78.052313}.
Nevertheless, interesting phenomena such as Entanglement
Sudden death \cite{barbosa2011disentanglement,Isar2011} and
persistence of entanglement even at infinite times
\cite{PhysRevA.105.042405} have been observed in such
studies. Tracking entanglement dynamics in two-mode states
is considerably simpler as computable measures for
entanglement such as logarithmic negativity and in some
cases the entanglement of formation can also be computed
analytically~\cite{PhysRevA.96.062338}.  While construction of
entanglement measures for multimode Gaussian states,
especially for pure states~\cite{PhysRevA.102.012421} has
been attempted, there is still no known computable measure
for general multimode Gaussian states. Entanglement witnesses for 
multi-mode Gaussian states
that have been
constructed~\cite{Hyllus_2006,Mihaescu_2020},
 are not useful for our purpose as we
strive to extract separability features and these witnesses
are coarse-grained one-way conditions sufficient only to
detect the presence of entanglement.
Therefore, we employ Semidefinite Programming
(SDP)~\cite{Tavakoli_2024} in our work where  we use two
different SDPs to check whether a given Gaussian state is
separable in some bipartition.  A method based on Linear
Matrix Inequalities presented recently by
Shan~\cite{Ma_2020} and an SDP extension for Gaussian states
similar to the Doherty-Parrilo-Spedliari (DPS) hierarchy are
the two methods that we use in our work
~\cite{Rajarama_Bhat_2017,Lepp_j_rvi_2024}.

We dynamically track the evolution of various initial
multimode Gaussian states and analyze separability across
different bipartitions after one, two, three or all
the modes are subjected to
environmental noise of varying 
durations of time. In particular, we
analyze the robustness of entanglement across various cuts
of the multimode states, where for a given strength of
environmental interactions we define robustness as the
minimum time of environmental interactions required to make
the initial state fully separable.  Next, by combining the
PPT criteria for Gaussian states and after performing a
separability check via the SDP-based techniques discussed
above, we are equipped to detect bound entanglement across
any cut of multimode Gaussian states. This also allows us to
track the robustness of bound entanglement for an initial
four-mode bound entangled state. We find that under the
influence of the environment, bound entanglement persists
for a finite amount of time, after which the state becomes
completely separable.

Further, we find a
situation, where a class of initial four-mode states (that
we refer to as generalized four-mode squeezed vacuum (gFMSV)
states), in the presence of noise becomes \emph{bound
entangled} across one or more bipartitions for a finite
window of time, before finally becoming fully separable.
This feature is rather rare in the case of discrete variable
systems~\cite{vogel}
making our discovery of a transient bound entangled state
interesting. The existence of a bound entangled phase for a
finite duration of time for initial gFMSV states raises an
important question: Can the transient bound entangled phase
also be observed for other initially entangled four-mode
states? To address this question, we considered other
exemplary four-mode states.  However, no bound entangled
phase was observed for any considered state.  The above
negation holds true when the initial four-mode states are
chosen randomly as well. This perhaps resonates with the
observation
of~\cite{horodecki2001boundentanglementcontinuousvariables}
that bound entanglement in CV systems is rare. Finally, we
try to argue what is special about the initial gFMSV states
that enables it to support a transient bound entangled
phase.

The paper's contents are organized as follows. We set the
stage with a brief primer on the prerequisites in Sec.
\ref{sec:pre}, containing a review of the basics of
Gaussian states and detection
schemes for Gaussian bound entanglement in Sec. \ref{sec:gaussianbasics},
and modelling noisy dynamics of
Gaussian states in Sec. \ref{sec:noisy}. In Sec.
\ref{sec:robustness}, we analyze the dynamics of Gaussian
entanglement under the influence of the considered noisy
dynamics. We identify a class of initial states for which we
observe the emergence of a transient bound entangled phase
before all the bipartitions of the state become separable in
Sec.  \ref{sec:transient}. We perform a robustness analysis
for random initial Gaussian states in Sec. \ref{sec:random},
and for initial bound entangled states in Sec.
\ref{sec:boundent}. Finally, we provide a conclusion in Sec.
\ref{sec:conclusion}.
\begin{figure}[h]
\centering
\includegraphics[scale=1]{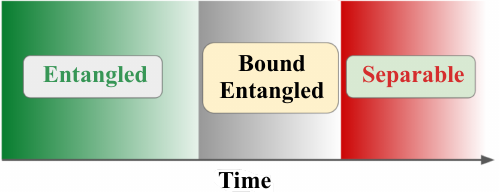}
\caption{Schematic of the time evolution of the entanglement
characteristics of an initial entangled state on interaction with a noisy
environment. }
\label{fig:enter-label}
\end{figure}
\section{Prerequisites}
\label{sec:pre}
We begin by providing a brief primer on the phase space
formalism for Gaussian states. Then we move on to describing
the semi-definite programs with details of the two methods
employed to detect Gaussian bound entanglement.  Finally,
the noise models under which the dynamics takes place are
discussed.
\subsection{Basics of Gaussian systems and Gaussian bound
entanglement}
\label{sec:gaussianbasics}
The $n$-mode CV systems are described
by the conjugate quadratures of
field operators, $\hat{\xi}$ denotes the
vector of the quadrature operators
$\hat{\xi} = (\hat{q}_1, \hat{p}_1, \hat{q}_2, \hat{p}_2,
\ldots, \hat{q}_n,
\hat{p}_n)^T$
 which satisfy the canonical commutation relations \[
[\hat{\xi}_\alpha, \hat{\xi}_\beta] = i\,\Omega_{\alpha\beta}
\quad\quad\quad
\alpha, \beta = 1, 2, \ldots, n
\]

 where $\Omega$ is the symplectic matrix:

\begin{eqnarray}
\Omega = \bigoplus_{i = 1}^n \omega
\quad\quad\quad
\omega =
\begin{pmatrix}
0 & 1 \\
-1 & 0
\end{pmatrix}.
\end{eqnarray}
The $2n \oplus 2n$ real symmetric covariance matrix of a
 an $n$-mode CV system
is defined through its elements $V_{\alpha\beta}=\langle\{
\Delta\hat{\xi}_\alpha, \Delta\hat{\xi}_\beta \}\rangle,$ where
$\Delta\hat{\xi}=\hat{\xi}_\alpha-\langle
\hat{\xi}_\alpha\rangle$ and
$\{\Delta\hat{\xi}_\alpha, \Delta\hat{\xi}_\beta \}=
\Delta\hat{\xi}_\alpha\Delta\hat{\xi}_\beta+\Delta\hat{\xi}_\beta\Delta\hat
{\xi}_\alpha.$\\
The uncertainty relation for any valid covariance matrix
$V$ translates to
\begin{equation}V + i\Omega \geq 0.
\label{eq:uncertainityreln}
\end{equation}
Eq. \eqref{eq:uncertainityreln} is a necessary and
sufficient condition for a real symmetric matrix $V$ to
correspond to a physical quantum state. The $n$-modes
can be partitioned in $\mathcal{N}(n) =\sum_{k=1}^{[n/2]}\binom{n}{k}$
ways, where $[x]$ denotes the integral part of $x$, and
$\binom{n}{m} = \frac{n!}{m!(n-m)!}$. The closed-form 
expression for $\mathcal{N}(n)$ is given by
\begin{eqnarray}
\mathcal{N}(n) =
\begin{cases} 
2^{n-1} - 1, & n ~\text{ is odd,} \\
2^{n-1} - 1 + {n!}/{2(\frac n2 !)^2}, & n ~\text{ is even.}
\end{cases}
\end{eqnarray}
The separability
criterion for an arbitrary $A:B$ partition of modes reads
as~\cite{PhysRevLett.86.3658}
\begin{equation}
V \geq V_{A} \oplus V_{B}
\end{equation}
According to a theorem by
Williamson~\cite{williamson1936algebraic}, every
positive-definite real symmetric matrix of even dimension
can be diagonalized through a symplectic transformation.
Therefore, given an arbitrary $n$-mode Gaussian state with
real symmetric covariance matrix, there exists a symplectic
matrix $S$ such that
\begin{align}
V=S[\bigoplus_{k=1}^{n}\nu_{k} I_{2}]S^{T},
\label{williamson}
\end{align}
where $S$ is a symplectic matrix ,{\it i.e.}
$S\Omega S^T = \Omega $.
The quantities $\nu_k$ are the symplectic eigenvalues
of $ V$ and they come in pairs.

Moreover, the symplectic matrix $S$ in
Equation~\eqref{williamson} can be decomposed using the
Euler decomposition. In fact, every $n$-mode symplectic
matrix $S$ can be written as
\begin{align}
S=K[\bigoplus_{k=1}^{n}S(r_{k})]L, 
\label{Sdecomp}
\end{align}
where $K$ and $L$ are orthogonal symplectic
matrices, and 
$S(r_{k})=\begin{pmatrix}
e^{-r_{k}} &0\\
0 &e^{r_{k}}
\end{pmatrix}$ 
is a set of single-mode squeezing matrices.
Combining Eq.~\eqref{williamson} and Eq.~\eqref{Sdecomp},
we obtain that an arbitrary $n$-mode covariance matrix
$ V $ can be written as
\begin{equation}
V = K[\oplus_{k=1}^n S(r_k)]L[\oplus_{k=1}^n\nu_k
I_2]L^T[\oplus_{k=1}^n S(r_k)]K^T\\
\label{eq:covgen}
\end{equation}

For an $n$-mode $(n \geq 2)$ Gaussian state, the partial
transpose with respect to a bipartition $A:B$ transforms the
covariance matrix 
\begin{equation}
V\stackrel{\rm PT}{\longrightarrow} \tilde{V} =
(\Omega_A\oplus \mathbb{I}_B)V(\Omega_A\oplus
\mathbb{I}_B)
\end{equation}
where $\Omega_A = \oplus_{k=1}^m \omega$
corresponds to a sign change of the momentum variables
belonging to subsystem $A$ consisting of modes $1$ to $m$.
The Positive under Partial Transpose (PPT) criterion can be
compactly expressed as
\begin{equation}
V + i\tilde{\Omega} \geq 0, ~~\text{with} ~~
\tilde{\Omega}
=\left(\begin{array}{cc}
-\Omega_{A} & 0 \\
0 & \Omega_{B} \\
\end{array} \right),
\label{eq:pptcriterion}
\end{equation}
where $\Omega_B = \oplus_{m+1}^N \omega$.
An entangled state that is PPT and thus not distillable is
called a bound entangled~\cite{Horodecki1997separability,
horodecki1998mixed}. For DV systems PPT criterion serves as
a necessary and sufficient condition of entanglement for $2
\times 2$ and $2 \times 3$
systems~\cite{Horodecki1996separability} and thus the
smallest system that supports bound entanglement is $2
\times 4$ \cite{Horodecki1997separability}.

For CV systems, especially for Gaussian
states, the characterization of entanglement using the PPT
criterion possesses a rich structure. For example, for any
$n$-mode Gaussian state, the PPT criterion provides a
necessary and sufficient criterion for separability for the
$1:(N-1)$ bipartition. Furthermore, PPT implies separability
for mono- and bi-symmetric Gaussian
states~\cite{Giedke_2001,Serafini_2005,Lami_2018}. For a
detailed account of Gaussian entanglement and the PPT
criterion see~\cite{Lami_2018}.

Establishing general criteria for the detection of arbitrary
PPT entangled states is quite difficult. However, there are
operational procedures for detecting entangled CV states
that have a positive partial
transpose~\cite{Ma_2020,Mihaescu_2020}. We review below the
two such techniques that we employ in our work.
\subsubsection{Detection using linear matrix inequalities (LMI)}
\label{sec:semidef1}
The method of constructing an entanglement witness for CV
systems was originally introduced by Hyllus and
Eisert~\cite{Hyllus_2006}. The Detection of Gaussian
Entanglement via Solving Linear Matrix inequalities was
developed by Ma {\it et. al.}~\cite{Ma_2020} and the method
has since been extended to detect entanglement of unknown CV
states via random measurements by Mihaescu {\it et.
al.}~\cite{Mihaescu_2020}.

For a Gaussian state with covariance matrix $V$,
our objective is to
find $V_{A} \,\&\, V_{B}$ such that
\begin{equation}
V -\left(\begin{array}{cc}
V_{A} & 0 \\
0 & V_{B}  \\
\end{array} \right)> \eta
\end{equation}
\begin{equation}
V -\left(\begin{array}{cc}
V_{A} & \Omega_{A} \\
\Omega_{A}^T & V_{A}  \\
\end{array} \right)> \eta
\end{equation}
\begin{equation}
V -\left(\begin{array}{cc}
V_{B} & \Omega_{B} \\
\Omega_{B}^T & V_{B}  \\
\end{array} \right)> \eta,
\end{equation}
where $\eta \geq 0$.
If we are able to find such $V_{A} \,\&\,
V_{B}$ the state with covariance matrix $V$ is
separable and conversely for all separable states such 
$V_{A} \,\&\,  V_{B}$ exist.
It should be mentioned that although the above Linear Matrix
inequalities are necessary and sufficient for checking the
separability of $V$, one has to be very careful when
the state $V$ lies very close to the boundary of
separable states or to the boundary of the set of physical
states ($i.e.$ the smallest eigenvalue of $V + i\Omega$
is very close to zero.) For such cases, since the
constraints are non-strict LMIs, the unavoidable round-off
errors caused by floating-point computations can have an
impact on the solvability of the problem. In these cases,
one solves an updated version of the problem where $\eta$ is
greater than some small negative number $\epsilon$.
For all practical purposes,
 we can choose $\eta>-\epsilon$
such that $10^{-6} < |\epsilon| <
10^{-9}$~\cite{Ma_2020}.
Specifically, for all calculations presented in this paper, we fix
$|\epsilon|=10^{-8}$.
\subsubsection{SDP via Symmetric extension of Gaussian States}
\label{sec:semidef2}
We can verify the entanglement of Gaussian states via
another SDP program which is based on the extension of
Doherty-Parrilo-Spedalieri (DPS) hierarchy to Gaussian
Systems. While complete extendibility was originally
discussed in \cite{Rajarama_Bhat_2017}, a proper framework
was developed by Lami {\it et.
al.}~\cite{PhysRevLett.123.050501} who introduced the
concept of $k$-extendibility of Gaussian States. They also
showed that the Wolfe-Werner Bound entangled state is not
2-extendible~\cite{PhysRevLett.123.050501}.

Since entanglement properties of Gaussian states only depend
on their covariance matrices, without any loss of
generality, we can restrict ourselves to Gaussian states
with zero mean. Hence a $(m+n)$-mode zero mean Gaussian
state in $\Gamma(\mathbb{C}^m) \otimes \Gamma(\mathbb{C}^n)$
is determined by a $2(m+n) \times 2(m+n)$ covariance matrix
\begin{equation}
V = \left(\begin{array}{cc}
A  & B \\
B^T& C \\
\end{array} \right)
\end{equation}
Hence if a Gaussian state $\rho$ is $k$-extendable with
respect to the second system, then there exists a real
matrix $\theta_k$ of order $2n \times 2n$ such that the
extended matrix
\begin{equation}
V_k = \left(\begin{array}{c|cccc}
A & B & B  & \cdots  & B \\
\hline
B^T & C &  \theta_k  & \cdots & \theta_k \\
B^T & \theta_k^T &    & \cdots & \theta_k \\
\vdots & \vdots &\vdots & \ddots & \vdots \\
B^T & \theta_k^T & \theta_k^T  & \cdots & C\end{array} \right)
\end{equation}
is the covariance matrix of a Gaussian state in
$\Gamma(\mathbb{C}^m) \otimes \Gamma(\mathbb{C}^n)^{\otimes
k}$

For a \( (n_A + n_B) \)-mode Gaussian state \( \rho_{AB} \)
with covariance matrix \( V_{AB} \), a necessary and
sufficient condition for \( k \)-extendibility can be
written as follows. There must exist a \( 2n_B \times 2n_B
\) covariance matrix \( \Delta_B \geq i \Omega_B \) such
that

\begin{eqnarray}
V_{AB} \geq i \Omega_A \oplus \left( \left( 1 - \frac{1}{k}
\right) \Delta_B
+ \frac{1}{k} i \Omega_B \right)
\end{eqnarray}
This equation for $k$-extendibility and complete
extendibility can be cast as an SDP program and can be used
to detect Gaussian bound entangled states
efficiently~\cite{Lepp_j_rvi_2024}.

Finally, by compiling everything together the task of
checking whether a given multi-mode Gaussian state is bound
entangled across any given bipartition reduces to a two-step
process
\begin{enumerate} 
\setlength{\itemindent}{-6pt}
\item
Consider the partial transposition and check its sign using
the criterion in Eq. \eqref{eq:pptcriterion}.
\item Check whether the state is separable across the
considered cut using
semidefinite programming techniques as laid out above.
\end{enumerate}
A given multi-mode Gaussian state is bound entangled
across a considered bipartition if it has positive partial
transposition (PPT) in that cut but it is inseparable across
the same cut.

Summing up, to check whether a Gaussian state 
is entangled across a given bipartition, we first apply the
PPT criterion. If we obtain a negative eigenvalue,
we conclude the state is entangled across that bipartition. 
However, if the state is PPT across the given bipartition and 
each partition contains at least two modes, it may be either 
separable or bound entangled. Therefore, in this case, we employ
the above mentioned SDP techniques to conclusively establish 
whether the state is separable or PPT (bound) entangled across
that bipartition.

\subsection{Modelling Noisy Dynamics}
\label{sec:noisy}

\begin{figure}[h]
\centering
\includegraphics[scale=1]{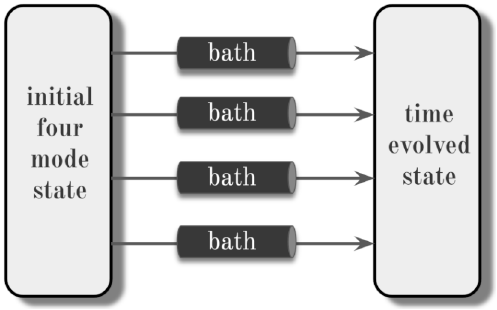}
\caption{Schematic of the local bath acting on four mode
states.}
\label{fig:bath}
\end{figure}

We consider an $n$-mode system interacting with a thermal
bath. The bath modes correspond to radiation modes which 
are comprised of a large number of harmonic oscillators. 
The corresponding total Hamiltonian can be written as
$$ \hat{H} = \hat{H}_{S}+ \hat{H}_{B} + \hat{H}_{S+B} $$ 
where $\hat{H}_{S}$ and
$\hat{H}_{B}$ are the system and the bath Hamiltonians
respectively, and $\hat{H}_{S+B}$ is the interaction
Hamiltonian.
\begin{equation}
 H = \sum_{i=1}^n \omega a_i^{\dagger}a_i +
\sum_{\textbf{k}}\omega_{\textbf{k}}
b_{\textbf{k}}^{\dagger}b_{\textbf{k}}
+
\sum_{i=1}^n (\Gamma a_i^{\dagger} + \Gamma^{\dagger}a_i),
\end{equation}
where the heat bath operators $\Gamma =
\sum_{\textbf{k}}g_{\textbf{k}}b_{\textbf{k}}$ and $\Gamma^{\dagger} =
\sum_{\textbf{k}}g_{\textbf{k}}^{*}b_{\textbf{k}}^{\dagger}$ with
$g_{\textbf{k}}$ the system-environment coupling and $a_i$,$b_{\textbf{k}}$ and
$a_i^{\dagger}$,$b_{\textbf{k}}^{\dagger}$ are annihilation
and creation
operators respectively.
Without loss of generality, the reservoir is assumed to be a
squeezed bath with
the following correlations~\cite{gardiner}:
\begin{eqnarray} \label{correlations}
\langle \Gamma^{\dagger}(t)\Gamma(t')\rangle &=& \gamma N
\delta(t-t'),
\nonumber \\
\langle \Gamma(t)\Gamma^{\dagger}(t')\rangle &=& \gamma (N+1)
\delta(t-t')\nonumber \\
\langle \Gamma(t)\Gamma(t')\rangle &=& \gamma M
\delta(t-t')\nonumber \\
\langle \Gamma^{\dagger}(t)\Gamma^{\dagger}(t')\rangle &=&
\gamma M^{*}
\delta(t-t'),
\end{eqnarray}
where \(\gamma\) is the damping rate, \(N\) represents the
mean photon number of
the squeezed reservoir, and \(M\) is a parameter related to
the phase
correlations of the squeezed reservoir. The Heisenberg
uncertainty relation
implies the constraint \(|M|^2 \leq N(N+1)\).
Under the Markovian assumption, we can write the master
equation for the reduced
density matrix of the $n$-mode field as
\begin{equation}\label{master_eqn_global}
\begin{aligned}
\frac{\partial }{\partial t} \rho =
\sum_{i,j=1}^{n} \Bigg[
& \frac{\gamma_i}{2} (N_i + 1) \left( 2a_i \rho
a_j^{\dagger} - a_i^{\dagger}
a_j \rho - \rho a_i^{\dagger} a_j \right) \\
&+ \frac{\gamma_i}{2} N_i \left( 2a_i^{\dagger} \rho a_j -
a_i a_j^{\dagger}
\rho - \rho a_i a_j^{\dagger} \right) \\
&+ \frac{\gamma_i}{2} M_i \left( 2a_i^{\dagger} \rho a_j^{\dagger} -
a_i^{\dagger} a_j^{\dagger} \rho - \rho a_i^{\dagger} a_j^{\dagger} \right)\\
&+ \frac{\gamma_i}{2} M_i^* \left( 2a_i \rho a_j - \rho a_i a_j - a_i a_j \rho
\right) \Bigg]
               \end{aligned}
\end{equation}
For the local bath considered in our analysis, the relevant
contributions come from the $i=j$ terms, while the cross
terms do not contribute.  Here, $N_i$ is the mean photon
number of the $i$th squeezed reservoir and $\gamma_i$ is the
damping rate of the $i$th mode. $M_i$ is a parameter related
to the phase correlations of the $i$th squeezed reservoir.
Moreover we work with a bath set at zero temperature.
For calculational simplicity, we set all $M_i = 0$ in the
rest of the paper. Therefore, our bath is a thermal bath. 
Moreover, we choose the damping rates of
all the bath modes to be identical, $i.e.$, $\gamma_i = \gamma$.
The bath is schematically described in Fig.~\ref{fig:bath}.

An $n$-mode state described by density matrix $\rho$ has the following
Weyl characteristic function :
\begin{equation}
\chi(\{\beta\}_n) = \chi (\beta_1,\beta_2,\dots,\beta_n) =
\text{Tr}[\rho
D(\beta_1,\beta_2,\dots,\beta_n)], \nonumber
\end{equation}
where $D(\beta_1,\beta_2,\dots,\beta_n) = \otimes_{k=1}^n
D_k(\beta_k)$ is the
$n$-mode displacement operator and $D_k(\beta_k) = \exp
(\beta_k a_k^{\dagger} -
\beta_k^* a_k)$
is the single-mode displacement operator.
Using the characteristic function, we can transform the
above master equation in the form of an equation for a
characteristic function. Then, through the relation $
\chi(\beta, t) = \exp\;\{-\frac{1}{2} \Lambda^T V(t)
\Lambda\}, $
where \(\Lambda = (\Lambda_1, \Lambda_2, \dots,
\Lambda_n)^{T} \in
\mathbb{R}^{2n}\) is a column vector,
we obtain a time-evolved expression purely in terms of
covariance matrix of the
initial state $V(0)$~\cite{Olivares_2012}
\begin{equation}\label{evolution_equation}
V(t) = X(t)V_1(0)X(t)^T + Y(t),
\end{equation}
where for four-mode states, we have
\begin{eqnarray}
X(t) = \frac14 (1-\tau)^{\frac{1}{4}}\; \mathbb{I}_8, \text{
and } Y(t) =
F^{\oplus 4},
\end{eqnarray}
where $\tau = 1 - e^{-2\gamma t}$ with $\tau = 0
\Leftrightarrow t=0$ and $\tau
= 1 \Leftrightarrow t=\infty$. We call $\tau$ to be the
regularized time to
contrast it with the physical time $t$. Here
 $F = (\frac{1}{2}+N) (1-\sqrt{1-\tau})\mathbb{I}_2$ and
$\mathbb{I}_k$ is the
$k \times k$ identity matrix.
\section{Dynamics of Gaussian Entanglement: robustness against noise}
\label{sec:robustness}
In this section, we examine the evolution of various
Gaussian states of the four-mode CV system in the presence
of environmental interactions modeled via a bath.
Interaction with a bath is expected to lead to decoherence
and to reduction in correlations thereby decreasing the
amount of entanglement present in the state with time. The
time up to which the entanglement is retained defines the
robustness of the chosen initial state under the influence
of noise. Formally, we define robustness in the following
way
\begin{definition}
The robustness of an initially entangled state to noise is
defined by the minimum time $\tau^*$ it takes for the bath
to make the initial state separable across all bipartitions,
\begin{eqnarray}
\tau^* = \min \tau, \nonumber \\
s.t. ~~\Lambda_{\tau^*}(\rho_{in}) \in \text{ SEP},
\end{eqnarray}
where $\Lambda_\tau$ is the dynamical map corresponding to
the noisy evolution and $\rho_{in}$ denotes the initial
state.  Here $\tau$ is connected to physical time via the
following relation $\tau = 1 - e^{-2 \gamma t}$, and $SEP$
denotes the set of state
separable across all partitions.  \label{def:def1}
\end{definition}

Our analysis reveals that although the general intuitive
picture where  an  initially entangled state finally
becoming separable under the influence of noise remains
true, in certain cases we find a transient phase where
the state become bound entangled. This on the one hand leads
to a new way to find bound entangled or PPT entangled
states within the family of Gaussian states and on the other hand,
it illustrates how distillable entanglement can transition into bound entanglement in the presence of noise. This adds to a new
and surprising feature to the evolution of entanglement
under a noisy environment.

Below we present our investigation of dynamics of various
Gaussian states in the presence of noise. We begin our
analysis by considering the noisy dynamics of Gaussian
states that are initially entangled.
\subsection{Generalized FMSV states: transient bound
entanglement}
\label{sec:transient}
\begin{figure}[h]
\centering
\includegraphics[scale=1]{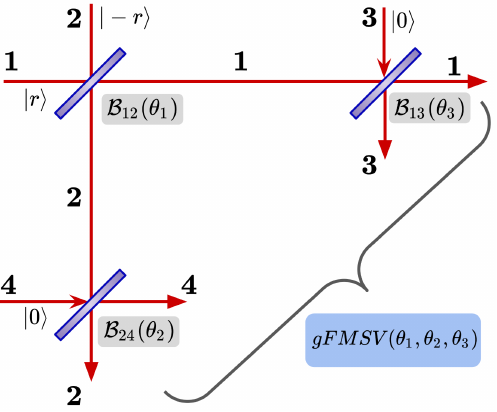}
\caption{Schematic of the optical setup for the
generation of the generalized four mode squeezed vacuum
(FMSV) states. The transmission coefficient of a beam
splitter parameterized by an angle $\theta_i$ is $\cos^2
\theta_i$. The standard FMSV state is obtained when all the
three beam splitters are balanced ones, i.e., $\theta_i =
\pi/4$ for $i = 1,2,$ and $3$.}
\label{fig:fmsv}
\end{figure}
Consider a three-parameter family of four mode Gaussian
states generated by the schematic optical setup shown in
Fig.~\ref{fig:fmsv}. We refer to this family as the
generalized four-mode squeezed vacuum (gFMSV) states. The
state parameters can be tuned by changing the
transmissivities of the three beam splitters used to
generate the gFMSV states. The covariance matrix of a gFMSV
states $V_{gFMSV}$ state can be expressed as
\begin{equation}
 V_{gFMSV} =  \overline{\mathcal{B}}_{13}(\theta_3)\overline{\mathcal{B}}_{24}(\theta_2)\overline{\mathcal{B}}_{12} \Gamma_r (\theta_1)\overline{\mathcal{B}}_{12}^T(\theta_1)\overline{\mathcal{B}}_{24}^T(\theta_2)\overline{\mathcal{B}}_{13}^T(\theta_3),
    \label{eq:gfmsv}
\end{equation}
where $\Gamma_r = $ diag$\{e^r,e^{-r}\} ~\oplus$
diag$\{e^{-r},e^r\}$ is the tensor product of two
single-mode squeezed states of squeezing $r$ and $-r$
respectively.  Here $\overline{\mathcal{B}}_{ij}(\theta) =
\mathcal{B}_{ij}(\theta) \oplus \mathbb{I}_k$ denotes the
beam splitter action on modes $i$ and $j$ while no action is
implemented on the remaining mode $k$, where we have $i\neq
j \neq k \in \{1,2,3\}$. The beam splitter operation can be
conveniently expressed as
\begin{eqnarray}
B_{ij}(\theta) = \begin{bmatrix}
\cos \theta & 0 & \sin \theta & 0 \\
0 & \cos \theta & 0 & \sin \theta \\
-\sin \theta & 0 & \cos \theta & 0 \\
0 & -\sin \theta & 0 & \cos \theta
\end{bmatrix},
\end{eqnarray}
where $\cos^2 \theta$ is the transmission coefficient of the
beam splitter.

In our work, we will primarily focus on a special 
case of the gFMSV state, referred to as just
the four-mode squeezed vacuum (FMSV) states where all the
beam splitters in Equation~\eqref{eq:gfmsv} are chosen to be
balanced, {\it i.e.}, $\theta_1=\theta_2 =\theta_3 = \pi/4$.  The
central result of our manuscript is the analysis of the
dynamics of this particular FMSV  states under the
influence of noisy 
environments.  Note that in general, 
four-mode squeezed vacuum states comprise 
entangled as well as separable states. However, in our work, when we
mention the ``FMSV state'' we refer to a specific state as
considered in~\cite{Ma1990,Das2016,PhysRevA.102.012421}
which is entangled across all bipartitions and
generated via the schematic described in Fig.~\ref{fig:fmsv}.
The FMSV is a genuinely entangled
state with the covariance matrix $V_{\text{FMSV}}$,
which is given by
\begin{equation*}
\begin{bmatrix}
\cosh^2 r ~\mathbb{I}_2 & \frac{1}{2}\sinh 2r ~\sigma_z &
\sinh^2 r
~\mathbb{I}_2 & \frac{1}{2}\sinh 2r ~\sigma_z \\
\frac{1}{2}\sinh 2r ~\sigma_z & \cosh^2 r ~\mathbb{I}_2 &
\frac{1}{2}\sinh 2r
~\sigma_z & \sinh^2 r ~\mathbb{I}_2 \\
\sinh^2 r ~\mathbb{I}_2 & \frac{1}{2}\sinh 2r ~\sigma_z &
\cosh^2 r
~\mathbb{I}_2 & \frac{1}{2}\sinh 2r ~\sigma_z \\
\frac{1}{2}\sinh 2r ~\sigma_z & \sinh^2 r ~\mathbb{I}_2 &
\frac{1}{2}\sinh 2r
~\sigma_z & \cosh^2 r ~\mathbb{I}_2
\end{bmatrix}
\end{equation*}
with \(\mathbb{I}_2\) and \(\sigma_z\) being the identity
and Pauli matrix in the $z$-direction respectively. The FMSV
state with a moderately low squeezing strength $r$ can be
prepared in the laboratories  by using linear optical
elements like $50$:$50$ beam splitters and two single-mode
squeezed vacuum states.

In the presence of local baths acting independently on all
the four modes, the time-evolved covariance matrix
can be obtained using Eq.~(\ref{evolution_equation}),
where $V_1(0)$ is taken as the initial FMSV covariance
matrix above and is given as the following
\begin{equation}
\label{fmsv_evolved}
V_{FMSV}(t) =
\left(\begin{array}{cc|cc}
a \mathbb{I}_2 & b \sigma_z & c \mathbb{I}_2 & b \sigma_z \\
b \sigma_z & a \mathbb{I}_2 & b \sigma_z & c \mathbb{I}_2 \\
\hline
c \mathbb{I}_2 & b \sigma_z & a \mathbb{I}_2 & b \sigma_z \\
b \sigma_z & c \mathbb{I}_2 & b \sigma_z & a \mathbb{I}_2
\end{array}\right),
\end{equation}
where
\begin{equation*}
\begin{aligned}
a &= (1/2+N)(1-\sqrt{1-\tau})+\cosh^2\:r \sqrt{1-\tau} \\
b &= (1/2)\sinh\:2r~(1-\tau)^{\frac{1}{2}} \\
c &= \sinh^2 \: r ~(1-\tau)^{\frac{1}{2}}\\
\tau &= 1 - e^{-2\gamma t}
\end{aligned}
\end{equation*}
The FMSV state under consideration has four modes and we can
consider situations where one, two, three or all of them
undergo noisy evolution and thereupon we can examine
entanglement across different partitions.  Since the initial
FMSV state is NPT, identifying the NPT phase of the evolved
state amounts to checking only PPT criterion. Once all the
symplectic eigenvalues of PPT criterion turn positive, we
employ SDP techniques to identify whether the state is
entangled or separable.

When the local noise acts on any one mode and we consider
separability in the $12$:$34$ partition, the state does not
disentangle after a finite amount of time. This is also the
case when the local noise acts on certain pairs of modes,
namely, the modes $\{1,4\}$ or $\{2,3\}$.  However, in the case
when noise acts on the pair of modes $\{1,2\}$ and $\{3,4\}$ the
entanglement survives upto a certain time and then
disappears. The time at which entanglement disappears
depends upon the mean photon number of the bath.  In the
case where any three of the undergo evolution under the
noisy environment the result are similar and the initial NPT
entangled FMSV states becomes separable after a certain time
which again depends upon the mean photon number of the bath.
Finally, if all the four modes are evolved under a noisy
environment the initial NPT entangled FMSV state looses
entanglement ever earlier.
We demonstrate these trends in Table~\ref{tab:fmsv_evolution}
for different noise strengths.
\begin{table}[h!]
\centering
\begin{tabular}{|>{\centering\arraybackslash}p{4.8cm}|
>{\centering\arraybackslash}p{1cm}|
>{\centering\arraybackslash}p{1cm}|
>{\centering\arraybackslash}p{1.1cm}|}
       \hline
       \textbf{Noisy Modes} & \multicolumn{3}{c|}{$\tau^*$} \\
       \hline
       & $N = 2$ & $N = 4$ & $N = 10$ \\
       \hline
       $\{1\}, \{2\}, \{3\}, \{4\}$ & - & - & - \\
       \hline
       $\{1,2\}, \{3,4\}$ & 0.82 & 0.60 & 0.32 \\
       $\{1,4\}, \{2,3\}$ & - & - & - \\
       \hline
       $\{1,2,3\}, \{1,2,4\}, \{1,3,4\}, \{2,3,4\}$ & 0.71 & 0.48 & 0.24 \\
       \hline
       $\{1,2,3,4\}$ & 0.38 & 0.20 & 0.09 \\
\hline
\end{tabular}
\caption{Robustness analysis of the FMSV state for $N = 2,
4, 10$ (mean photon number of the bath) and $r = 0.6$
(squeezing parameter). The table is divided into four parts
based on the number the of modes on which the noise acts.
For example \texttt{\{$i$\}} implies that the noise is
applied on a single mode $i$,similarly \texttt{\{$1,2$\}}
implies that noise is being applied to both modes $1$ and
$2$, and so on. Clearly, $\tau^*$ decreases with increasing
noise strength $N$.}
\label{tab:fmsv_evolution}
\end{table}
\subsubsection{Transient bound entanglement phase}
A closer analysis of time evolved FMSV state when the local
noise acts on two next to next neighboring modes reveals a
striking feature that there is a transient bound entangled
phase before the entanglement disappears.  In this case, the
time-evolved FMSV state after being NPT entangled for some
time becomes PPT entangled at least in one bipartition.
This introduces a new time scale $\tau_{\text{BE}}$ in the
problem over and above $\tau^*$ (see
Definition~\ref{def:def1}). After being bound entangled for
a finite temporal window $[\tau_{\text{BE}}, \tau^*)$, the
state ultimately becomes fully separable at $\tau = \tau^*$.
Therefore, in this case, the dynamics can be split into
three distinct temporal regions

\begin{enumerate}
\setlength{\itemindent}{-0.6pt}
\item $\tau \in [0,\tau_{\text{BE}})$: the time evolved
state is NPT entangled at least across one bipartition.
\item $\tau \in [\tau_{\text{BE}},\tau^*)$: the time evolved
state is PPT
(bound) entangled at least across one bipartition and is
separable across the
other cuts.
\item $\tau \in [\tau^*,1)$: the time evolved state is
separable across all
bipartitions.
\end{enumerate}
This counterintuitive phase of bound entanglement
occurs when the
next to next modes such as $\{1,3\}$ or $ \{2,4\}$ are
subjected to a local noisy environment.
The  two time scales 
$\tau_{\text{BE}}$ and  $\tau^*)$ for different average
photon number are tabulated in
Table~\ref{tab:fmsv_boundent}.
\begin{table}[h!]
   \centering
   \begin{tabular}{|>{\centering\arraybackslash}p{2cm}|
                   >{\centering\arraybackslash}p{2cm}|
                   >{\centering\arraybackslash}p{1.5cm}|
                   >{\centering\arraybackslash}p{1.5cm}|}

       \hline
 \textbf{Noisy Modes} & \textbf{$N$} & $\tau_{BE}$ & $\tau^*$ \\
 \hline
 \multirow{3}{*}{$\{1,3\}, \{2,4\}$} & 2 & 0.71 & 0.82 \\
 & 4 & 0.48 & 0.60 \\
 & 10 & 0.24 & 0.32 \\
 \hline
 \end{tabular}
\caption{Comparison of $\tau_{BE}$ and $\tau^*$ for noisy
modes $\{1,3\}$ or $\{2,4\}$ of an initial FMSV states with
squeezing parameter $r = 0.6$ for various mean photon
numbers of the bath, $N = 2, 4, 10$. We observe that the
transient bound entangled phase is quite rigid and retains
itself for strong noise values as well. As expected, with
increasing noise strength (mean photon number of the bath),
the duration for which the bound entangled phase survives,
reduces.}
\label{tab:fmsv_boundent}
\end{table}

It is important to pin down the partitions across which the
transient bound entanglement phase is present. Firstly, the
only possibility of getting a PPT entangled state for a
$4-$mode Gaussian state is in the $2:2$ bipartitions. A
$4-$mode Gaussian state has three $2:2$ bipartitions,
namely, $$\{[12 : 34], \, [13 : 24],
\, [14 : 23]\}.$$ It turns out that the time-evolved FMSV
state is bound entangled in the partition $\{[14 : 23]\}$ in
the same way as in $\{[12 : 34]\}$. However, in the
partition $\{ [13 : 24]\}$ there is no phase of bound
entanglement, see Table~\ref{tab:bipartitions}.  In
contrast, one should note that the Wolfe-Werner state (one of the
first examples of Gaussian bound entangled state) is
bound entangled only in the $\{[12 : 34]\}$ bipartition
while it is separable in the bipartitions $\{ \, [13 : 24],
\, [14 : 23]\}$~\cite{10661902}. We shall discuss more on the
robustness properties of the Wolfe-Werner state in
subsequent sections.

\begin{table}[h!]
\centering
\begin{tabular}{|>{\centering\arraybackslash}p{2cm}|
 >{\centering\arraybackslash}p{2cm}|
 >{\centering\arraybackslash}p{1.5cm}|
 >{\centering\arraybackslash}p{1.5cm}|}
\hline
\textbf{Noisy Modes} & \textbf{Bipartition} & $\tau_{BE}$ & $\tau^*$ \\
\hline
\multirow{3}{*}{$\{1,3\}, \{2,4\}$} & $12 : 34$ & 0.48 & 0.60 \\
  & $14 : 23$ & 0.48 & 0.60 \\
  & $13 : 24$ & - & 0.60 \\
  \hline
\end{tabular}
\caption{Summary of bipartitions and corresponding
$\tau_{BE}$ and $\tau^*$ values for \(N = 4\) (mean photon
number) of an initial FMSV state with squeezing parameter
\(r = 0.6\).}
\label{tab:bipartitions}
\end{table}

Such a transient bound entangled phase is observed for a
broad class of gFMSV states defined in Eq.~\eqref{eq:gfmsv}.
The exact ranges of beam splitter transmissivities $\cos^2
\theta_i$s defining this class depend on the squeezing
strength of the initial single-mode squeezed state $r$, the
decay rate $\gamma$, and the average number of photons in
the bath $N$. See Table.  \ref{tab:angles_table} for some
cases for which an initial gFMSV state supports a transient
bound entangled phase during dynamics.
\begin{table}[h!]
   \centering
   \begin{tabular}{|>{\centering\arraybackslash}p{1.8cm}|
                   >{\centering\arraybackslash}p{2.3cm}|
                   >{\centering\arraybackslash}p{2cm}|
                   >{\centering\arraybackslash}p{0.85cm}|
                   >{\centering\arraybackslash}p{0.85cm}|}
       \hline
       \textbf{Noisy Modes} & \textbf{Pattern of \(\theta_i\)s} &
\textbf{Choice of the variable angle} & $\tau_{BE}$ & $\tau^*$ \\
       \hline
       \multirow{5}{*}{$\{1,3\}, \{2,4\}$}
       & \multirow{7}{*}{\shortstack{$\theta_1 = \theta_2 = \theta_3 $ \\ $ =
\theta_i$ }}
       & $30^\circ = \pi/6$ & - & - \\
       & & {$39^\circ$} & - & 0.85 \\
       & & $40^\circ$ & 0.77 & 0.78 \\
       & & $44^\circ$ & 0.52 & 0.61 \\
       & & $45^\circ = \pi/4$ & 0.48 & 0.60 \\
       \hline
       \multirow{5}{*}{$\{1,3\}, \{2,4\}$}
       & \multirow{7}{*}{\shortstack{$\theta_1 $ (variable)\\ $\theta_2 =
\theta_3 = \pi/4$ }}
       & $30^\circ = \pi/6$ & - & - \\
       & & {$39^\circ$} & - & 0.85 \\
       & & $40^\circ$ & 0.76 & 0.78 \\
       & & $44^\circ$ & 0.52 & 0.61 \\
       & & $45^\circ = \pi/4$ & 0.48 & 0.60 \\
       \hline
       \multirow{7}{*}{$\{1,3\}, \{2,4\}$}
       & \multirow{7}{*}{\shortstack{$\theta_1 = \pi/4$ \\ $\theta_2 =
\theta_3$ (variable)}}
       & $9^\circ$ & - & 0.60 \\
       & & $10^\circ$ & 0.59 & 0.60 \\
       & & $15^\circ = \pi/12$ & 0.59 & 0.60 \\
       & & $30^\circ = \pi/6$ & 0.55 & 0.60 \\
       & & $40^\circ$ & 0.58 & 0.60 \\
       & & $44^\circ$ & 0.48 & 0.60 \\
       & & $45^\circ = \pi/4$ & 0.48 & 0.60 \\
       \hline
   \end{tabular}
\caption{Analysis of $\tau_{BE}$ and $\tau^*$ for the bound
entangled phase of the initial gFMSV states under the
variation of beamsplitter angles (transmissivities). In
particular we consider three specific cases (i) $\theta_1 =
\theta_2 = \theta_3 = \theta_i$. We note that in this case
the bound entangled phase vanishes at angle value of 39
degrees. (ii) $\theta_1 $ (variable), $\theta_2 = \theta_3 =
\pi/4$. This case quite similar to the case (i) as we note
that bound entangled vanishes at the same angle value.(iii)
$\theta_1 = \pi/4$ , $\theta_2 = \theta_3$ (variable) This
is the most interesting case as it turns out that such a
configuration can sustain a transient bound entangled phase
for a very large angle value. Due to the symmetry of the
state we note that the values for $\tau_{BE}$ and $\tau^*$
in the range $45^\circ - 90^\circ$ are same as $0^\circ -
45^\circ$. Here we have set \(N = 4\) (mean photon number)
and \(r = 0.6\) (squeezing parameter) of both the initial
single mode squeezed states, see Fig.~\ref{fig:fmsv}.}
\label{tab:angles_table}
\end{table}

Our analysis reveals that whenever $\theta_1$ is varying the
transient bound entangled phase is relatively fragile and
vanishes after a small angle (transmissivity) change on the
subsequent beam splitter actions. However, when $\theta_1$
is kept unchanged and fixed to the balanced configuration of
$\theta_1 = \pi/4$, and $\theta_2$ and $\theta_3$ is varied,
the transient bound entangled phase is sustained for
considerably larger variations in
subsequent beam splitter angle values. This points to the
strong role of the initial beam splitter that entangles
modes $1$ and $2$. If it is chosen to be balanced, the final
output state tends to be more robust in sustaining a transient 
bound entangled phase under noisy dynamics, even with variations
in the transmissivities of the subsequent beam splitters.
Table~\ref{tab:angles_table} highlights instances of this
feature. Further, this analysis also demonstrates that
the bound entangled transient phase exists for a range of
parameters values of the gFMSV family and is not a singular
phenomena.

\subsection{Other Gaussian
states}
\label{sec:exemplary}
We continue our investigation by taking other typical
four-mode Gaussian states as initial
states before they are subjected to noisy dynamics. However,
unlike the case of the FMSV state, we
do not find a transient bound entangled phase in all those
cases. States that are initially entangled transition directly to
separable states after interacting with the bath.  Specifically,
we present results for two such states.

A four-mode Gaussian state constructed by the tensor product of
two two-mode squeezed vacuum (TMSV) states with the same
squeezing strength $r$. The covariance matrix of the state
read as
\begin{eqnarray}
V_{TMSV^{\otimes 2}} = V_{TMSV}^{13}\oplus
V_{TMSV}^{24},
\end{eqnarray}
where the subscript ${ij}$ denote modes $i$ and $j$
respectively, and
\begin{eqnarray}
V_{TMSV} = \begin{bmatrix}
\cosh 2r \mathbb{I}_{2} & \sinh 2r \sigma_z \\
\sinh 2r \sigma_z &   \cosh 2r \mathbb{I}_{2}
\end{bmatrix}.
\end{eqnarray}
As is the case with FMSV state, The state
$V_{TMSV^{\otimes 2}}$ is initially NPT and we can
verify entanglement via the PPT criterion, once state
evolves and the symplectic eigenvalues turn positive, we
employ SDP to check whether the state is PPT entangled or
separable. It is found that the state either remains
entangled or becomes separable across all bipartitions
after interacting with the bath for a given time depending
upon the modes that interact with the bath. Therefore, there
exists only one time scale in this problem, namely the
robustness time $\tau^*$. We present the robustness analysis
of this state for different cases where one, two, three or
all modes of interact with local noisy baths in
Table~\ref{tab:combined_fmsv2}.
\begin{table}[h!]
\centering
\begin{tabular}{|>{\centering\arraybackslash}p{4.8cm}|
>{\centering\arraybackslash}p{1cm}|
>{\centering\arraybackslash}p{1cm}|
>{\centering\arraybackslash}p{1.1cm}|}
\hline
\textbf{Noisy Modes} & \multicolumn{3}{c|}{$\tau^*$} \\
\hline
 & $N = 2$ & $N = 4$ & $N = 10$ \\
 \hline
 $\{1\}, \{2\}, \{3\}, \{4\}$ & - & - & - \\
 \hline
 $\{1,2\}, \{1,4\}, \{2,3\}, \{3,4\}$ & 0.82 & 0.60 & 0.32 \\
 \hline
 $\{1,3\}, \{2,4\}$ & - & - & - \\
 \hline
 $\{1,2,3\}, \{1,2,4\}, \{1,3,4\}, \{2,3,4\}$ & 0.82 & 0.60 & 0.32 \\
 \hline
 $\{1,2,3,4\}$ & 0.54 & 0.31 & 0.14 \\
 \hline
\end{tabular}
\caption{Comparison of $\tau^*$ values for noise acting on
different mode
configurations for varied strength of the bath \(N=2, 4,
10\) (mean photon
numbers), where the squeezing parameter of the TMSV states
are fixed to $r =
0.6$.}
\label{tab:combined_fmsv2}
\end{table}

The second example we choose is from a work by Adesso
{\it et.~al.}~\cite{Adesso_2007}.

We begin with an uncorrelated four-mode state, where 
each mode is initially in the vacuum state of its respective
Fock space. The corresponding covariance matrix (CM) is the
identity matrix. We apply a two-mode squeezing transformation 
with squeezing parameter \( s \) to modes 2 and 3, followed
by two additional two-mode squeezing transformations with 
squeezing parameter \( a \) to the pairs of modes (1,2) and (3,4).
These transformations redistribute the initial pairwise entanglement
across all four modes. For any values of \( s \) and \( a \), 
the output is a pure four-mode Gaussian state with CM \(
V\), given by:
\begin{equation}
V = \mathcal{S}_{3,4}(a) \mathcal{S}_{1,2}(a)
\mathcal{S}_{2,3}(s) \mathcal{S}_{2,3}^\mathsf{T}(s)
\mathcal{S}_{1,2}^\mathsf{T}(a)
\mathcal{S}_{3,4}^\mathsf{T}(a),
\end{equation}
where \( \mathcal{S}_{i,j}(r) \) is the two-mode squeezing matrix defined as:

\begin{equation}
\mathcal{S}_{i,j}(r) =
\begin{pmatrix}
\cosh r & \sinh r \\
\sinh r & \cosh r
\end{pmatrix}
\oplus
\begin{pmatrix}
\cosh r & -\sinh r \\
-\sinh r & \cosh r
\end{pmatrix}.
\end{equation}
One should note that due to this particular construction we
have : (i) \( \mathcal{S}_{i,j} = \mathcal{S}_{j,i} \), and
(ii) symplectic operations on disjoint mode pairs commute.
Consequently, the covariance matrix
 remains invariant under simultaneous
exchange of modes \( 1 \leftrightarrow 4 \) and \( 2
\leftrightarrow 3 \).

The Covariance Matrix \( V \) has the following block structure:
\begin{equation}\label{adesso_state}
V = \left(\begin{array}{cc|cc}
\sigma_1 & \varepsilon_{12} & \varepsilon_{13}  &\varepsilon_{14}\\
\varepsilon_{12} & \sigma_2 & \varepsilon_{23} & \varepsilon_{24} \\
\hline
\varepsilon^T_{13} & \varepsilon^T_{23}  & \sigma_3 &\varepsilon_{34}\\
\varepsilon^T_{14} & \varepsilon^T_{24}  & \varepsilon^T_{34} & \sigma_4 \\
\end{array} \right),
\end{equation}

where the diagonal blocks \( \sigma_i \) and off-diagonal blocks \( \varepsilon_{ij} \) are given by:

\begin{align}
    \sigma_1 = \sigma_4 &= \left[\cosh^2(a) + \cosh(2s) \sinh^2(a)\right] \mathbf{I}_2, \notag \\
    \sigma_2 = \sigma_3 &= \left[\cosh(2s) \cosh^2(a) + \sinh^2(a)\right] \mathbf{I}_2, \notag \\
    \varepsilon_{1,2} = \varepsilon_{3,4} &= \left[\cosh^2(s) \sinh(2a)\right] \mathbf{Z}_2, \notag \\
    \varepsilon_{1,3} = \varepsilon_{2,4} &= \left[\cosh(a) \sinh(a) \sinh(2s)\right] \mathbf{I}_2, \notag \\
    \varepsilon_{1,4} &= \left[\sinh^2(a) \sinh(2s)\right] \mathbf{Z}_2, \notag \\
    \varepsilon_{2,3} &= \left[\cosh^2(a) \sinh(2s)\right] \mathbf{Z}_2.
\end{align}

Here, \( \mathbf{I}_2 \) and \( \mathbf{Z}_2 \) denote the \( 2 \times 2 \)
identity and Pauli $\sigma_z$ matrices, respectively.

The state is quite interesting since it is fully inseparable
(\textit{i.e.} it contains genuine four-partite entanglement) for all
values of $s$ and $a$. Moreover, in the limit of very high
$a$, the state reproduces the entanglement value of two
EPR-like pairs (1,2 and 3,4). We find that this
state too does not
display a phase of bound entanglement for all the cases where
one, two, three or all the modes are allowed to interact
with a local bath. The analysis of the robustness of
entanglement of this state for various cases is provided in the
Table~\ref{tab:fmsv3_combined}.
\begin{table}[h!]
   \centering
   \begin{tabular}{|>{\centering\arraybackslash}p{3cm}|
                   >{\centering\arraybackslash}p{1.5cm}|
                   >{\centering\arraybackslash}p{1.5cm}|
                   >{\centering\arraybackslash}p{1.5cm}|}
       \hline
       \textbf{Noisy Modes} & \multicolumn{3}{c|}{$\tau^*$} \\
       \hline
       & \(N = 2\) & \(N = 4\) & \(N = 10\) \\
       \hline
       $\{1\}, \{4\}$ & - & - & - \\
       $\{2\}$ & 0.82 & 0.61 & 0.32 \\
       $\{3\}$ & 0.85 & 0.60 & 0.32 \\
       \hline
       $\{1,2\}, \{3,4\}$ & 0.82 & 0.61 & 0.32 \\
       $\{1,3\}, \{2,4\}$ & 0.64 & 0.41 & 0.20 \\
       $\{1,4\}$ & - & - & - \\
       $\{2,3\}$ & 0.47 & 0.26 & 0.11 \\
       \hline
       $\{1,2,3\}, \{2,3,4\}$ & 0.44 & 0.24 & 0.11 \\
       $\{1,2,4\}, \{1,3,4\}$ & 0.62 & 0.40 & 0.20 \\
       \hline
       $\{1,2,3,4\}$ & 0.41 & 0.23 & 0.10 \\
       \hline
   \end{tabular}
\caption{Comparison of $\tau^*$ values for the state whose covariance matrix
is expressed in Eq. \ref{adesso_state} for various mean photon numbers and
state parameters fixed to \(s = a = 0.6\).}
   \label{tab:fmsv3_combined}
\end{table}

In the subsequent section, we will search for the existence
of other classes of states that may support an intermediate 
bound entangled state. 
\section{Randomly chosen initial states: pure vs mixed}
\label{sec:random}
Having found the transient bound entangled state during
the evolution of four mode Gaussian NPT states it is
important to find out how ubiquitous is this phenomena  
($i.e.$ how common is
the emergence of a transient bound entangled phase for
different choices of initial states). We
consider Haar uniformly generated random pure and mixed
Gaussian states as initial states, evolve them under
the noisy environment and look for bound entanglement.

\subsection{Random pure four-mode Gaussian states}
We choose the initial state to be a random
pure four-mode Gaussian state with a fixed average energy.
From Eq.~\eqref{eq:covgen}, we have a general form of the 
covariance matrix $V = O \Gamma O^T,$
where $O$ is
an orthogonal symplectic matrix and $\Gamma $ is the tensor
product of $n$ single-mode squeezed states~\cite{Simon1994}.
Note that the
group of orthogonal symplectic matrices $K(n)$ is isomorphic
to the unitary group $U(n)$.  Using this isomorphism, one
can Haar uniformly generate an orthogonal symplectic matrix
using the following relation
\begin{equation}
U \in U(n) \rightarrow \begin{bmatrix}
\text{Re}(U) & \text{Im}(U)\\
-\text{Im}(U) & \text{Re}(U)
\end{bmatrix} = O(U)
\end{equation}
The energy constraint $$\text{Tr} ~V = \text{Tr}
~\Gamma = E,$$ fixes the choice of the single mode squeezers
composing $\Gamma$.  
For more details on generating Haar uniformly
generating random pure Gaussian states
see~\cite{roytypical2024}.

We considered $10^4$ random four-mode
Gaussian pure states generated by the aforementioned
formalism as the initial states. For all such states, 
irrespective of the energy bound $E$, we find that the time-evolved state
does not support a transient bound entangled phase for any
of the $2:2$ bipartitions, which are $\{[12 : 34], \, [13 :
24], \, [14 : 23]\}$.  The above result cements the
uniqueness of the gFMSV states and reinforces the idea that
bound entanglement for continuous variable systems is a rare
phenomenon~\cite{horodecki2001boundentanglementcontinuousvariables}.

The reason for the unique feature of the gFMSV states may be
argued from the unique structure of
the symplectic spectrum of the gFMSV class of states
not present in the random states.
Specifically, we focus on the spectrum of the matrix used to
check the PPT criterion for a given covariance matrix, {\it
i.e.}, $V + i\tilde{\Omega}$.  The spectrum of $V
+ i\tilde{\Omega}$ for gFMSV states reveals four
eigenvalues, each with a multiplicity of $2$. In the
non-positive-transpose (NPT) phase, two of these eigenvalues
are negative (with identical magnitudes, $\lambda_1 =
\lambda_2 < 0$). As expected, these negative eigenvalues
transition to positive values as the system enters the bound
entangled phase. Consequently, bound entangled gFMSV states
retain the same structured spectrum\---four eigenvalues,
each with multiplicity 2\---but with all eigenvalues now
positive ($\lambda_i > 0\ \forall\, i$).  Hence, bound
entangled states arising from gFMSV states exhibit a
characteristic pattern of four eigenvalues, each with a
multiplicity of two.  In contrast, states generated via the
Haar uniform procedure exhibit a random structure in the
eigenvalues of $V + i\tilde{\Omega}$. Moreover, random mixed
NPT four-mode Gaussian states generated following the
procedure in \cite{Lepp_j_rvi_2024} not only lack a 
structured eigenvalue pattern but also typically possess
only a single negative eigenvalue, unlike the two 
negative eigenvalues observed in gFMSV states.
\subsection{Random mixed four-mode Gaussian states}
Next, we consider the initial state of the
dynamics to be a random mixed Gaussian state. Rather than
generating Haar-uniform mixed states by tracing out modes
from Haar-random pure states, we follow the direct
generation method proposed in~\cite{Lepp_j_rvi_2024}. This
approach constructs random Gaussian covariance matrices by
modifying elements of the Gaussian Orthogonal Ensemble (GOE)
to ensure they satisfy condition (\ref{eq:uncertainityreln}). 
 Specifically, for a given GOE matrix \( G \), the
corresponding random quantum covariance matrix is defined as
\begin{equation}
V_{random} = G + \lambda_{\max}(i \Omega_{2n} - G) I_{2n},
\end{equation}
where the shift guarantees positive-definiteness.

The primary motivation for this method stems from two key
observations: (i) the resulting ensemble of random mixed
Gaussian covariance matrices is invariant under the action
of the ortho-symplectic group \( K(2n) = Sp(2n) \cap O(2n)
\), ensuring a natural and well-defined probability
distribution, and (ii) each element of the GOE is mapped to
the closest valid Gaussian covariance matrix in the operator
norm, providing a mathematically rigorous and efficient way
to generate states.  For a detailed discussion on the
generation and properties of these covariance matrices,
see~\cite{Lepp_j_rvi_2024}. Since our focus is restricted to
four-mode states, we employ this technique specifically to
generate four-mode mixed Gaussian states.

We generate $100$ random NPT states and check whether they
support any intermediate bound entangled state during their
evolution. Our analysis reveals that in this case, also, we
do not observe any transient bound entangled state, the
random states directly become fully separable from being
entangled after interacting with the bath.

\begin{table}[h]
\centering
\begin{tabular}{|>{\centering\arraybackslash}p{4cm}|
                >{\centering\arraybackslash}p{2cm}|}
    \hline
   \textbf{Noisy Modes} & $\tau^*$ \\
   \hline
   $\{i\}$ & 0.42 \\
   \hline
   $\{i, i+1\}$ & 0.14 \\
   $\{i, i+2\}$ & 0.16 \\
   $\{i, i+3\}$ & 0.17 \\
   \hline
   $\{i, i+1, i+2\}$ & 0.08 \\
   $\{i, i+1, i+3\}$ & 0.08 \\
   $\{i, i+2, i+3\}$ & 0.09 \\
   \hline
   $\{1,2,3,4\}$ & 0.06 \\
   \hline
   \end{tabular}
\caption{Average robustness time $\tau^*$ for $100$ random
states generated by modifying the elements of the Gaussian
orthogonal ensemble (GOE) for \(N = 4\) (mean photon number)
of the bath. }
\label{tab:randomgoe}
\end{table}

On a different note, to get an idea of how robust these
states are to noise, we compute the Haar averaged value of
the robustness time $\tau^*$ for typical choices of system
and noise parameters. The various Haar averaged values of
$\tau^*$ for noises acting on different modes are depicted
in Table~\ref{tab:randomgoe}.
\section{Robustness Analysis of Bound Entangled States}
\label{sec:boundent}
We take up here the robustness analysis of already
known four mode Gaussian bound entangled states.
The first example of a bound entangled Gaussian state was
found by Werner and Wolf~\cite{PhysRevLett.86.3658} is a four mode state with
covariance matrix
\begin{equation}
\label{wolfewerner}
V_{WW} =\left(\begin{array}{cccccccc}
2 & 0  &  0 &  0 & 1 & 0 &  0 & 0\\
0 & 1  &  0 &  0 & 0 & 0 &  0 &-1\\
0 & 0  &  2 &  0 & 0 & 0 & -1 & 0\\
0 & 0  &  0 &  1 & 0 &-1 &  0 & 0\\
1 & 0  &  0 &  0 & 2 & 0 &  0 & 0\\
0 & 0  &  0 & -1 & 0 & 4 &  0 & 0\\
0 & 0  & -1 &  0 & 0 & 0 &  2 & 0\\
0 & -1 &  0 &  0 & 0 & 0 &  0 & 4
\end{array} \right)
\end{equation}
The eigenvalues of $V_{WW} + i\tilde{\Omega}$ are: $0$,
$3-\sqrt{3}$, $3$, $3+\sqrt{3}$ each with multiplicity $2$.
Incidentally, we note that the eigenvalues of gFMSV states
$V_{gFMSV} + i\tilde{\Omega}$ supporting bound
entangled phase also have a similar structure.  Later
multiple states of this form have been found~\cite{Ma_2019}.
These states can be generalized via covariance matrices of
the following form
\begin{equation}
V_{{\tt BE}} = \left(\begin{array}{cc}
\mathcal{A} &\mathcal{C} \\
\mathcal{C} &\mathcal{B} \\
\end{array} \right),
\end{equation}
where $\mathcal{A} = $ 
diag $\{A,B,A,B\}$, $\mathcal{B} = $
diag $\{C,D,C,D\}$, and

\begin{equation}\label{werner_c}
\mathcal{C} = \left(\begin{array}{cccc}
E &0 & 0 &0\\
0& 0 & 0 & -F \\
0 & 0 & -F &0\\
0 & -F & 0 & 0 \\
\end{array} \right).
\end{equation}
Such a state is known as the generalized Werner-Wolf state
and the necessary criterion of separability of the
generalized Werner-Wolf state is given
below~\cite{PhysRevA.107.022410}:
\begin{equation}
(AC-E^2)(BD-F^2)-2|EF|-CD-AB+1 \geq 0.
\end{equation}

In our analysis we evolve the Werner-Wolf state under
local noisy bath environment and consider various cases
where one, two, three or all the modes interact with the
environment. 
The entanglement analysis of the evolved state is
then carried 
out using its covariance matrix, which is obtained by substituting
 Equation (\ref{wolfewerner}) as the initial covariance matrix into 
Equation (\ref{evolution_equation}). The resulting covariance matrix
retains a generalized structure.
\begin{equation}  
V_{{\tt BE}}(t) = \left(\begin{array}{cc}  
\mathcal{A'} & \mathcal{C'} \\  
\mathcal{C'} & \mathcal{B'}  
\end{array}\right),  
\end{equation}  
with the blocks defined as:  
- Diagonal matrices:  
  \begin{align*}  
  \mathcal{A'} &= \text{diag}\{A', B', A', B'\}, \\  
  \mathcal{B'} &= \text{diag}\{C', D', C', D'\},  
  \end{align*}  
  where  
  \begin{align*}  
  A' = C' &= \frac{1}{2} + N + \left(\frac{3}{2} - N\right)\sqrt{1 - \tau}, \\  
  B' &= \frac{1}{2} + N - \left(N - \frac{1}{2}\right)\sqrt{1 - \tau}, \\  
  D' &= \frac{1}{2} + N + \left(\frac{7}{2} - N\right)\sqrt{1 - \tau}.  
  \end{align*}  
and $\mathcal{C'}$ is of the same form as eq (\ref{werner_c}) with $E = F = \sqrt{1 - \tau}$

As the initial state is Bound entangled we proceed with SDP 
techniques provided above in to detect 
entanglement/separability of the evolved state.
Although we present the robustness analysis for the
specific case of the Werner-Wolf state,
the results remain qualitatively similar when one considers
the generalized versions.
The bound entanglement of Werner-wolf state
turns out to be robust and lasts for a regularized time from
$0$ to $\tau^*$. The results are summarized in
Table~\ref{tab:combined}. We show the $\tau^*$ values for
for noise in different modes and for different average
photon number in the bath modes. Even under strong noise 
conditions, the bound entanglement persists for a finite
duration before eventually transitioning to a separable 
state across all bipartitions.

The results are summarized in
Table~\ref{tab:combined}, where we present the $\tau^*$
values for noise in different modes and for varying average
photon numbers in the bath modes. The bound entanglement
remains resilient within the regularized time range $\tau
\in [0, \tau^*)$, even when the initial state is subjected
to strong noise. However, if the interaction with the bath
modes continues, the state eventually becomes separable
across all bipartitions.
\begin{table}[h!]
\centering
\begin{tabular}{|>{\centering\arraybackslash}p{2.5cm}|
>{\centering\arraybackslash}p{1.5cm}|
>{\centering\arraybackslash}p{1.5cm}|
>{\centering\arraybackslash}p{1.5cm}|}
\hline
\textbf{Noisy Modes} & \multicolumn{3}{c|}{$\tau^*$} \\
\hline
& $N = 2$ & $N = 4$ & $N = 10$ \\
\hline
$\{1\}, \{2\}, \{3\}, \{4\}$ & 0.82 & 0.60 & 0.32 \\
\hline
$\{1,2\}$ & 0.15 & 0.07 & 0.03 \\
$\{1,3\}, \{2,4\}$ & 0.37 & 0.19 & 0.08 \\
$\{1,4\}, \{2,3\}$ & 0.32 & 0.17 & 0.07 \\
$\{3,4\}$ & 0.47 & 0.26 & 0.11 \\
\hline
$\{1,2,3\}, \{1,2,4\}$ & 0.13 & 0.06 & 0.03 \\
$\{1,3,4\}, \{2,3,4\}$ & 0.24 & 0.12 & 0.05 \\
\hline
$\{1,2,3,4\}$ & 0.12 & 0.06 & 0.03 \\
\hline
\end{tabular}
\caption{Values of $\tau^*$ for different modes and mean
photon numbers \(N = 2, 4, 10\) for the Werner-Wolf bound
entangled state. Looking at the table one notes that when
we induce the bath in a single mode, we see that the state
becomes separable after a fairly long phase of bound
entanglement. Moving on to the case two of modes, we notice
that the states become separable more readily since we are
inducing dissipation in two modes. The pattern continues
with the case of three modes and finally, we see that when
inducing noise in all modes, the bound entanglement is lost
quite quickly.}
\label{tab:combined}
\end{table}

\section{Conclusion}
\label{sec:conclusion}
In this work, we tracked the entanglement dynamics of
initially entangled four-mode Gaussian states when they
evolve under the influence of a Markovian noise modeled by a
bath of harmonic oscillators. The bath modes were made to
interact with one, two, three or all the four modes of the
system.  Our approach based on SDP allowed us to capture the
exact moment at which the state becomes separable. The time
marking the onset of separability was used to characterize
the robustness of a wide variety of entangled four mode
Gaussian states and also Gaussian bound entangled states.

Secondly, and perhaps the most striking piece of our result
is that for a large class of generalized four-mode squeezed
vacuum (gFMSV) states when chosen as the initial state of
the dynamics, we observe an \textit{intermediate bound
entangled phase} for a finite time window. Of course,
finally, the state succumbs to noise and becomes separable
across all bipartitions.  This feature seems robust since
the intermediate bound entangled phase is observed for a
wide range of system parameters and the phase lasts for a
considerable amount of time. Interestingly, such a phase was
not observed for any other choice of initial states, not
even when the initial state was chosen randomly from a Haar
uniform ensemble.
Therefore, on the one hand, for almost all choices of initial
states, we do not observe any transient bound entangled phase,
as has been pointed out earlier. On the other hand, we identify
a family of states characterized by three continuous parameters
that exhibit this transient bound entangled phase. This 
observation aligns with the rarity of bound entanglement in 
CV states, as previously noted
in~\cite{horodecki2001boundentanglementcontinuousvariables}.

While we have limited our analysis to four-mode Gaussian
states, one can immediately note that our framework can be
easily generalized to study environmental noise in
other multimode
systems. In future work, we plan to investigate the
evolution of four mode states in more general noise models
such as Non-Markovian environments with different spectral
densities~\cite{refId0,An2007} and explore the possibility
of having transient bound entangled phase.


\begin{thebibliography}{66}%
\makeatletter
\providecommand \@ifxundefined [1]{%
 \@ifx{#1\undefined}
}%
\providecommand \@ifnum [1]{%
 \ifnum #1\expandafter \@firstoftwo
 \else \expandafter \@secondoftwo
 \fi
}%
\providecommand \@ifx [1]{%
 \ifx #1\expandafter \@firstoftwo
 \else \expandafter \@secondoftwo
 \fi
}%
\providecommand \natexlab [1]{#1}%
\providecommand \enquote  [1]{``#1''}%
\providecommand \bibnamefont  [1]{#1}%
\providecommand \bibfnamefont [1]{#1}%
\providecommand \citenamefont [1]{#1}%
\providecommand \href@noop [0]{\@secondoftwo}%
\providecommand \href [0]{\begingroup \@sanitize@url \@href}%
\providecommand \@href[1]{\@@startlink{#1}\@@href}%
\providecommand \@@href[1]{\endgroup#1\@@endlink}%
\providecommand \@sanitize@url [0]{\catcode `\\12\catcode `\$12\catcode
  `\&12\catcode `\#12\catcode `\^12\catcode `\_12\catcode `\%12\relax}%
\providecommand \@@startlink[1]{}%
\providecommand \@@endlink[0]{}%
\providecommand \url  [0]{\begingroup\@sanitize@url \@url }%
\providecommand \@url [1]{\endgroup\@href {#1}{\urlprefix }}%
\providecommand \urlprefix  [0]{URL }%
\providecommand \Eprint [0]{\href }%
\providecommand \doibase [0]{http://dx.doi.org/}%
\providecommand \selectlanguage [0]{\@gobble}%
\providecommand \bibinfo  [0]{\@secondoftwo}%
\providecommand \bibfield  [0]{\@secondoftwo}%
\providecommand \translation [1]{[#1]}%
\providecommand \BibitemOpen [0]{}%
\providecommand \bibitemStop [0]{}%
\providecommand \bibitemNoStop [0]{.\EOS\space}%
\providecommand \EOS [0]{\spacefactor3000\relax}%
\providecommand \BibitemShut  [1]{\csname bibitem#1\endcsname}%
\let\auto@bib@innerbib\@empty
\bibitem [{\citenamefont {Weedbrook}\ \emph {et~al.}(2012)\citenamefont
  {Weedbrook}, \citenamefont {Pirandola}, \citenamefont {García-Patrón},
  \citenamefont {Cerf}, \citenamefont {Ralph}, \citenamefont {Shapiro},\ and\
  \citenamefont {Lloyd}}]{Weedbrook_2012}%
  \BibitemOpen
  \bibfield  {author} {\bibinfo {author} {\bibfnamefont {C.}~\bibnamefont
  {Weedbrook}}, \bibinfo {author} {\bibfnamefont {S.}~\bibnamefont
  {Pirandola}}, \bibinfo {author} {\bibfnamefont {R.}~\bibnamefont
  {García-Patrón}}, \bibinfo {author} {\bibfnamefont {N.~J.}\ \bibnamefont
  {Cerf}}, \bibinfo {author} {\bibfnamefont {T.~C.}\ \bibnamefont {Ralph}},
  \bibinfo {author} {\bibfnamefont {J.~H.}\ \bibnamefont {Shapiro}}, \ and\
  \bibinfo {author} {\bibfnamefont {S.}~\bibnamefont {Lloyd}},\ }\href
  {\doibase 10.1103/revmodphys.84.621} {\bibfield  {journal} {\bibinfo
  {journal} {Reviews of Modern Physics}\ }\textbf {\bibinfo {volume} {84}},\
  \bibinfo {pages} {621–669} (\bibinfo {year} {2012})}\BibitemShut {NoStop}%
\bibitem [{\citenamefont {Adesso}\ \emph {et~al.}(2014)\citenamefont {Adesso},
  \citenamefont {Ragy},\ and\ \citenamefont {Lee}}]{Adesso_2014}%
  \BibitemOpen
  \bibfield  {author} {\bibinfo {author} {\bibfnamefont {G.}~\bibnamefont
  {Adesso}}, \bibinfo {author} {\bibfnamefont {S.}~\bibnamefont {Ragy}}, \ and\
  \bibinfo {author} {\bibfnamefont {A.~R.}\ \bibnamefont {Lee}},\ }\href
  {\doibase 10.1142/s1230161214400010} {\bibfield  {journal} {\bibinfo
  {journal} {Open Systems \& Information Dynamics}\ }\textbf {\bibinfo {volume}
  {21}},\ \bibinfo {pages} {1440001} (\bibinfo {year} {2014})}\BibitemShut
  {NoStop}%
\bibitem [{\citenamefont {Arvind}\ \emph
  {et~al.}(1995{\natexlab{a}})\citenamefont {Arvind}, \citenamefont {Dutta},
  \citenamefont {Mukunda},\ and\ \citenamefont {Simon}}]{Arvind_1995}%
  \BibitemOpen
  \bibfield  {author} {\bibinfo {author} {\bibnamefont {Arvind}}, \bibinfo
  {author} {\bibfnamefont {B.}~\bibnamefont {Dutta}}, \bibinfo {author}
  {\bibfnamefont {N.}~\bibnamefont {Mukunda}}, \ and\ \bibinfo {author}
  {\bibfnamefont {R.}~\bibnamefont {Simon}},\ }\href {\doibase
  10.1007/bf02848172} {\bibfield  {journal} {\bibinfo  {journal} {Pramana}\
  }\textbf {\bibinfo {volume} {45}},\ \bibinfo {pages} {471–497} (\bibinfo
  {year} {1995}{\natexlab{a}})}\BibitemShut {NoStop}%
\bibitem [{\citenamefont {Arvind}\ \emph
  {et~al.}(1995{\natexlab{b}})\citenamefont {Arvind}, \citenamefont {Dutta},
  \citenamefont {Mukunda},\ and\ \citenamefont {Simon}}]{arvind_two_mode}%
  \BibitemOpen
  \bibfield  {author} {\bibinfo {author} {\bibnamefont {Arvind}}, \bibinfo
  {author} {\bibfnamefont {B.}~\bibnamefont {Dutta}}, \bibinfo {author}
  {\bibfnamefont {N.}~\bibnamefont {Mukunda}}, \ and\ \bibinfo {author}
  {\bibfnamefont {R.}~\bibnamefont {Simon}},\ }\href {\doibase
  10.1103/PhysRevA.52.1609} {\bibfield  {journal} {\bibinfo  {journal} {Phys.
  Rev. A}\ }\textbf {\bibinfo {volume} {52}},\ \bibinfo {pages} {1609}
  (\bibinfo {year} {1995}{\natexlab{b}})}\BibitemShut {NoStop}%
\bibitem [{\citenamefont {Duan}\ \emph {et~al.}(2000)\citenamefont {Duan},
  \citenamefont {Giedke}, \citenamefont {Cirac},\ and\ \citenamefont
  {Zoller}}]{PhysRevLett.84.2722}%
  \BibitemOpen
  \bibfield  {author} {\bibinfo {author} {\bibfnamefont {L.-M.}\ \bibnamefont
  {Duan}}, \bibinfo {author} {\bibfnamefont {G.}~\bibnamefont {Giedke}},
  \bibinfo {author} {\bibfnamefont {J.~I.}\ \bibnamefont {Cirac}}, \ and\
  \bibinfo {author} {\bibfnamefont {P.}~\bibnamefont {Zoller}},\ }\href
  {\doibase 10.1103/PhysRevLett.84.2722} {\bibfield  {journal} {\bibinfo
  {journal} {Phys. Rev. Lett.}\ }\textbf {\bibinfo {volume} {84}},\ \bibinfo
  {pages} {2722} (\bibinfo {year} {2000})}\BibitemShut {NoStop}%
\bibitem [{\citenamefont {Simon}(2000)}]{PhysRevLett.84.2726}%
  \BibitemOpen
  \bibfield  {author} {\bibinfo {author} {\bibfnamefont {R.}~\bibnamefont
  {Simon}},\ }\href {\doibase 10.1103/PhysRevLett.84.2726} {\bibfield
  {journal} {\bibinfo  {journal} {Phys. Rev. Lett.}\ }\textbf {\bibinfo
  {volume} {84}},\ \bibinfo {pages} {2726} (\bibinfo {year}
  {2000})}\BibitemShut {NoStop}%
\bibitem [{\citenamefont {Arvind}\ and\ \citenamefont
  {Mukunda}(1999)}]{arvind_bell_cv1}%
  \BibitemOpen
  \bibfield  {author} {\bibinfo {author} {\bibnamefont {Arvind}}\ and\ \bibinfo
  {author} {\bibfnamefont {N.}~\bibnamefont {Mukunda}},\ }\href {\doibase
  https://doi.org/10.1016/S0375-9601(99)00471-5} {\bibfield  {journal}
  {\bibinfo  {journal} {Physics Letters A}\ }\textbf {\bibinfo {volume}
  {259}},\ \bibinfo {pages} {421} (\bibinfo {year} {1999})}\BibitemShut
  {NoStop}%
\bibitem [{\citenamefont {Kumar}\ \emph {et~al.}(2021)\citenamefont {Kumar},
  \citenamefont {Saxena},\ and\ \citenamefont {Arvind}}]{arvind_bell_cv2}%
  \BibitemOpen
  \bibfield  {author} {\bibinfo {author} {\bibfnamefont {C.}~\bibnamefont
  {Kumar}}, \bibinfo {author} {\bibfnamefont {G.}~\bibnamefont {Saxena}}, \
  and\ \bibinfo {author} {\bibnamefont {Arvind}},\ }\href {\doibase
  10.1103/PhysRevA.103.042224} {\bibfield  {journal} {\bibinfo  {journal}
  {Phys. Rev. A}\ }\textbf {\bibinfo {volume} {103}},\ \bibinfo {pages}
  {042224} (\bibinfo {year} {2021})}\BibitemShut {NoStop}%
\bibitem [{\citenamefont {Braunstein}\ and\ \citenamefont
  {Kimble}(2000)}]{cvdc1}%
  \BibitemOpen
  \bibfield  {author} {\bibinfo {author} {\bibfnamefont {S.~L.}\ \bibnamefont
  {Braunstein}}\ and\ \bibinfo {author} {\bibfnamefont {H.~J.}\ \bibnamefont
  {Kimble}},\ }\href {\doibase 10.1103/PhysRevA.61.042302} {\bibfield
  {journal} {\bibinfo  {journal} {Phys. Rev. A}\ }\textbf {\bibinfo {volume}
  {61}},\ \bibinfo {pages} {042302} (\bibinfo {year} {2000})}\BibitemShut
  {NoStop}%
\bibitem [{\citenamefont {Ralph}\ and\ \citenamefont
  {Huntington}(2002)}]{cvdc2}%
  \BibitemOpen
  \bibfield  {author} {\bibinfo {author} {\bibfnamefont {T.~C.}\ \bibnamefont
  {Ralph}}\ and\ \bibinfo {author} {\bibfnamefont {E.~H.}\ \bibnamefont
  {Huntington}},\ }\href {\doibase 10.1103/PhysRevA.66.042321} {\bibfield
  {journal} {\bibinfo  {journal} {Phys. Rev. A}\ }\textbf {\bibinfo {volume}
  {66}},\ \bibinfo {pages} {042321} (\bibinfo {year} {2002})}\BibitemShut
  {NoStop}%
\bibitem [{\citenamefont {Braunstein}\ and\ \citenamefont
  {Kimble}(1998)}]{cvtel1}%
  \BibitemOpen
  \bibfield  {author} {\bibinfo {author} {\bibfnamefont {S.~L.}\ \bibnamefont
  {Braunstein}}\ and\ \bibinfo {author} {\bibfnamefont {H.~J.}\ \bibnamefont
  {Kimble}},\ }\href {\doibase 10.1103/PhysRevLett.80.869} {\bibfield
  {journal} {\bibinfo  {journal} {Phys. Rev. Lett.}\ }\textbf {\bibinfo
  {volume} {80}},\ \bibinfo {pages} {869} (\bibinfo {year} {1998})}\BibitemShut
  {NoStop}%
\bibitem [{\citenamefont {Pirandola}\ and\ \citenamefont
  {Mancini}(2006)}]{Pirandola2006}%
  \BibitemOpen
  \bibfield  {author} {\bibinfo {author} {\bibfnamefont {S.}~\bibnamefont
  {Pirandola}}\ and\ \bibinfo {author} {\bibfnamefont {S.}~\bibnamefont
  {Mancini}},\ }\href {\doibase 10.1134/s1054660x06100057} {\bibfield
  {journal} {\bibinfo  {journal} {Laser Physics}\ }\textbf {\bibinfo {volume}
  {16}},\ \bibinfo {pages} {1418–1438} (\bibinfo {year} {2006})}\BibitemShut
  {NoStop}%
\bibitem [{\citenamefont {Patra}\ \emph {et~al.}(2022)\citenamefont {Patra},
  \citenamefont {Gupta}, \citenamefont {Roy},\ and\ \citenamefont
  {Sen(De)}}]{cvtel3}%
  \BibitemOpen
  \bibfield  {author} {\bibinfo {author} {\bibfnamefont {A.}~\bibnamefont
  {Patra}}, \bibinfo {author} {\bibfnamefont {R.}~\bibnamefont {Gupta}},
  \bibinfo {author} {\bibfnamefont {S.}~\bibnamefont {Roy}}, \ and\ \bibinfo
  {author} {\bibfnamefont {A.}~\bibnamefont {Sen(De)}},\ }\href {\doibase
  10.1103/PhysRevA.106.022433} {\bibfield  {journal} {\bibinfo  {journal}
  {Phys. Rev. A}\ }\textbf {\bibinfo {volume} {106}},\ \bibinfo {pages}
  {022433} (\bibinfo {year} {2022})}\BibitemShut {NoStop}%
\bibitem [{\citenamefont {Lami}\ \emph {et~al.}(2018)\citenamefont {Lami},
  \citenamefont {Serafini},\ and\ \citenamefont {Adesso}}]{Lami_2018}%
  \BibitemOpen
  \bibfield  {author} {\bibinfo {author} {\bibfnamefont {L.}~\bibnamefont
  {Lami}}, \bibinfo {author} {\bibfnamefont {A.}~\bibnamefont {Serafini}}, \
  and\ \bibinfo {author} {\bibfnamefont {G.}~\bibnamefont {Adesso}},\ }\href
  {\doibase 10.1088/1367-2630/aaa654} {\bibfield  {journal} {\bibinfo
  {journal} {New Journal of Physics}\ }\textbf {\bibinfo {volume} {20}},\
  \bibinfo {pages} {023030} (\bibinfo {year} {2018})}\BibitemShut {NoStop}%
\bibitem [{\citenamefont {Giorda}\ and\ \citenamefont {Paris}(2010)}]{gqd}%
  \BibitemOpen
  \bibfield  {author} {\bibinfo {author} {\bibfnamefont {P.}~\bibnamefont
  {Giorda}}\ and\ \bibinfo {author} {\bibfnamefont {M.~G.~A.}\ \bibnamefont
  {Paris}},\ }\href {\doibase 10.1103/PhysRevLett.105.020503} {\bibfield
  {journal} {\bibinfo  {journal} {Phys. Rev. Lett.}\ }\textbf {\bibinfo
  {volume} {105}},\ \bibinfo {pages} {020503} (\bibinfo {year}
  {2010})}\BibitemShut {NoStop}%
\bibitem [{\citenamefont {Lau}\ \emph {et~al.}(2017)\citenamefont {Lau},
  \citenamefont {Pooser}, \citenamefont {Siopsis},\ and\ \citenamefont
  {Weedbrook}}]{Lau2017}%
  \BibitemOpen
  \bibfield  {author} {\bibinfo {author} {\bibfnamefont {H.-K.}\ \bibnamefont
  {Lau}}, \bibinfo {author} {\bibfnamefont {R.}~\bibnamefont {Pooser}},
  \bibinfo {author} {\bibfnamefont {G.}~\bibnamefont {Siopsis}}, \ and\
  \bibinfo {author} {\bibfnamefont {C.}~\bibnamefont {Weedbrook}},\ }\href
  {\doibase 10.1103/PhysRevLett.118.080501} {\bibfield  {journal} {\bibinfo
  {journal} {Phys. Rev. Lett.}\ }\textbf {\bibinfo {volume} {118}},\ \bibinfo
  {pages} {080501} (\bibinfo {year} {2017})}\BibitemShut {NoStop}%
\bibitem [{\citenamefont {Das}\ \emph {et~al.}(2018)\citenamefont {Das},
  \citenamefont {Siopsis},\ and\ \citenamefont {Weedbrook}}]{weed}%
  \BibitemOpen
  \bibfield  {author} {\bibinfo {author} {\bibfnamefont {S.}~\bibnamefont
  {Das}}, \bibinfo {author} {\bibfnamefont {G.}~\bibnamefont {Siopsis}}, \ and\
  \bibinfo {author} {\bibfnamefont {C.}~\bibnamefont {Weedbrook}},\ }\href
  {\doibase 10.1103/PhysRevA.97.022315} {\bibfield  {journal} {\bibinfo
  {journal} {Phys. Rev. A}\ }\textbf {\bibinfo {volume} {97}},\ \bibinfo
  {pages} {022315} (\bibinfo {year} {2018})}\BibitemShut {NoStop}%
\bibitem [{\citenamefont {Killoran}\ \emph {et~al.}(2019)\citenamefont
  {Killoran}, \citenamefont {Bromley}, \citenamefont {Arrazola}, \citenamefont
  {Schuld}, \citenamefont {Quesada},\ and\ \citenamefont {Lloyd}}]{seth2019}%
  \BibitemOpen
  \bibfield  {author} {\bibinfo {author} {\bibfnamefont {N.}~\bibnamefont
  {Killoran}}, \bibinfo {author} {\bibfnamefont {T.~R.}\ \bibnamefont
  {Bromley}}, \bibinfo {author} {\bibfnamefont {J.~M.}\ \bibnamefont
  {Arrazola}}, \bibinfo {author} {\bibfnamefont {M.}~\bibnamefont {Schuld}},
  \bibinfo {author} {\bibfnamefont {N.}~\bibnamefont {Quesada}}, \ and\
  \bibinfo {author} {\bibfnamefont {S.}~\bibnamefont {Lloyd}},\ }\href
  {\doibase 10.1103/PhysRevResearch.1.033063} {\bibfield  {journal} {\bibinfo
  {journal} {Phys. Rev. Res.}\ }\textbf {\bibinfo {volume} {1}},\ \bibinfo
  {pages} {033063} (\bibinfo {year} {2019})}\BibitemShut {NoStop}%
\bibitem [{\citenamefont {Sharma}\ \emph {et~al.}(2025)\citenamefont {Sharma},
  \citenamefont {Kumar}, \citenamefont {Arora},\ and\ \citenamefont
  {Arvind}}]{sharma2025}%
  \BibitemOpen
  \bibfield  {author} {\bibinfo {author} {\bibfnamefont {M.}~\bibnamefont
  {Sharma}}, \bibinfo {author} {\bibfnamefont {C.}~\bibnamefont {Kumar}},
  \bibinfo {author} {\bibfnamefont {S.}~\bibnamefont {Arora}}, \ and\ \bibinfo
  {author} {\bibnamefont {Arvind}},\ }\href {https://arxiv.org/abs/2502.17182}
  {\enquote {\bibinfo {title} {Continuous variable quantum teleportation,
  $u(2)$ invariant squeezing and non-gaussian resource states},}\ } (\bibinfo
  {year} {2025}),\ \Eprint {http://arxiv.org/abs/2502.17182} {arXiv:2502.17182
  [quant-ph]} \BibitemShut {NoStop}%
\bibitem [{\citenamefont {Laudenbach}\ \emph {et~al.}(2018)\citenamefont
  {Laudenbach}, \citenamefont {Pacher}, \citenamefont {Fung}, \citenamefont
  {Poppe}, \citenamefont {Peev}, \citenamefont {Schrenk}, \citenamefont
  {Hentschel}, \citenamefont {Walther},\ and\ \citenamefont
  {H\"{u}bel}}]{Laudenbach2018}%
  \BibitemOpen
  \bibfield  {author} {\bibinfo {author} {\bibfnamefont {F.}~\bibnamefont
  {Laudenbach}}, \bibinfo {author} {\bibfnamefont {C.}~\bibnamefont {Pacher}},
  \bibinfo {author} {\bibfnamefont {C.~F.}\ \bibnamefont {Fung}}, \bibinfo
  {author} {\bibfnamefont {A.}~\bibnamefont {Poppe}}, \bibinfo {author}
  {\bibfnamefont {M.}~\bibnamefont {Peev}}, \bibinfo {author} {\bibfnamefont
  {B.}~\bibnamefont {Schrenk}}, \bibinfo {author} {\bibfnamefont
  {M.}~\bibnamefont {Hentschel}}, \bibinfo {author} {\bibfnamefont
  {P.}~\bibnamefont {Walther}}, \ and\ \bibinfo {author} {\bibfnamefont
  {H.}~\bibnamefont {H\"{u}bel}},\ }\href {\doibase 10.1002/qute.201800011}
  {\bibfield  {journal} {\bibinfo  {journal} {Advanced Quantum Technologies}\
  }\textbf {\bibinfo {volume} {1}} (\bibinfo {year} {2018}),\
  10.1002/qute.201800011}\BibitemShut {NoStop}%
\bibitem [{\citenamefont {Pirandola}\ and\ \citenamefont
  {Papanastasiou}(2024)}]{Pirandola2024}%
  \BibitemOpen
  \bibfield  {author} {\bibinfo {author} {\bibfnamefont {S.}~\bibnamefont
  {Pirandola}}\ and\ \bibinfo {author} {\bibfnamefont {P.}~\bibnamefont
  {Papanastasiou}},\ }\href {\doibase 10.1103/PhysRevResearch.6.023321}
  {\bibfield  {journal} {\bibinfo  {journal} {Phys. Rev. Res.}\ }\textbf
  {\bibinfo {volume} {6}},\ \bibinfo {pages} {023321} (\bibinfo {year}
  {2024})}\BibitemShut {NoStop}%
\bibitem [{\citenamefont {Zhang}\ \emph {et~al.}(2024)\citenamefont {Zhang},
  \citenamefont {Bian}, \citenamefont {Li}, \citenamefont {Yu},\ and\
  \citenamefont {Guo}}]{Zhang_2024}%
  \BibitemOpen
  \bibfield  {author} {\bibinfo {author} {\bibfnamefont {Y.}~\bibnamefont
  {Zhang}}, \bibinfo {author} {\bibfnamefont {Y.}~\bibnamefont {Bian}},
  \bibinfo {author} {\bibfnamefont {Z.}~\bibnamefont {Li}}, \bibinfo {author}
  {\bibfnamefont {S.}~\bibnamefont {Yu}}, \ and\ \bibinfo {author}
  {\bibfnamefont {H.}~\bibnamefont {Guo}},\ }\href {\doibase 10.1063/5.0179566}
  {\bibfield  {journal} {\bibinfo  {journal} {Applied Physics Reviews}\
  }\textbf {\bibinfo {volume} {11}} (\bibinfo {year} {2024}),\
  10.1063/5.0179566}\BibitemShut {NoStop}%
\bibitem [{\citenamefont {Braunstein}\ and\ \citenamefont {van
  Loock}(2005)}]{exp}%
  \BibitemOpen
  \bibfield  {author} {\bibinfo {author} {\bibfnamefont {S.~L.}\ \bibnamefont
  {Braunstein}}\ and\ \bibinfo {author} {\bibfnamefont {P.}~\bibnamefont {van
  Loock}},\ }\href {\doibase 10.1103/RevModPhys.77.513} {\bibfield  {journal}
  {\bibinfo  {journal} {Rev. Mod. Phys.}\ }\textbf {\bibinfo {volume} {77}},\
  \bibinfo {pages} {513} (\bibinfo {year} {2005})}\BibitemShut {NoStop}%
\bibitem [{\citenamefont {Horodecki}\ and\ \citenamefont
  {Lewenstein}(2000)}]{Horodecki_2000}%
  \BibitemOpen
  \bibfield  {author} {\bibinfo {author} {\bibfnamefont {P.}~\bibnamefont
  {Horodecki}}\ and\ \bibinfo {author} {\bibfnamefont {M.}~\bibnamefont
  {Lewenstein}},\ }\href {\doibase 10.1103/physrevlett.85.2657} {\bibfield
  {journal} {\bibinfo  {journal} {Physical Review Letters}\ }\textbf {\bibinfo
  {volume} {85}},\ \bibinfo {pages} {2657} (\bibinfo {year}
  {2000})}\BibitemShut {NoStop}%
\bibitem [{\citenamefont {Horodecki}\ \emph {et~al.}(2001)\citenamefont
  {Horodecki}, \citenamefont {Cirac},\ and\ \citenamefont
  {Lewenstein}}]{horodecki2001boundentanglementcontinuousvariables}%
  \BibitemOpen
  \bibfield  {author} {\bibinfo {author} {\bibfnamefont {P.}~\bibnamefont
  {Horodecki}}, \bibinfo {author} {\bibfnamefont {J.~I.}\ \bibnamefont
  {Cirac}}, \ and\ \bibinfo {author} {\bibfnamefont {M.}~\bibnamefont
  {Lewenstein}},\ }\href {https://arxiv.org/abs/quant-ph/0103076} {\enquote
  {\bibinfo {title} {Bound entanglement for continuous variables is a rare
  phenomenon},}\ } (\bibinfo {year} {2001}),\ \Eprint
  {http://arxiv.org/abs/quant-ph/0103076} {arXiv:quant-ph/0103076 [quant-ph]}
  \BibitemShut {NoStop}%
\bibitem [{\citenamefont {Werner}\ and\ \citenamefont
  {Wolf}(2001)}]{PhysRevLett.86.3658}%
  \BibitemOpen
  \bibfield  {author} {\bibinfo {author} {\bibfnamefont {R.~F.}\ \bibnamefont
  {Werner}}\ and\ \bibinfo {author} {\bibfnamefont {M.~M.}\ \bibnamefont
  {Wolf}},\ }\href {\doibase 10.1103/PhysRevLett.86.3658} {\bibfield  {journal}
  {\bibinfo  {journal} {Phys. Rev. Lett.}\ }\textbf {\bibinfo {volume} {86}},\
  \bibinfo {pages} {3658} (\bibinfo {year} {2001})}\BibitemShut {NoStop}%
\bibitem [{\citenamefont {Shor}\ \emph {et~al.}(2003)\citenamefont {Shor},
  \citenamefont {Smolin},\ and\ \citenamefont
  {Thapliyal}}]{PhysRevLett.90.107901}%
  \BibitemOpen
  \bibfield  {author} {\bibinfo {author} {\bibfnamefont {P.~W.}\ \bibnamefont
  {Shor}}, \bibinfo {author} {\bibfnamefont {J.~A.}\ \bibnamefont {Smolin}}, \
  and\ \bibinfo {author} {\bibfnamefont {A.~V.}\ \bibnamefont {Thapliyal}},\
  }\href {\doibase 10.1103/PhysRevLett.90.107901} {\bibfield  {journal}
  {\bibinfo  {journal} {Phys. Rev. Lett.}\ }\textbf {\bibinfo {volume} {90}},\
  \bibinfo {pages} {107901} (\bibinfo {year} {2003})}\BibitemShut {NoStop}%
\bibitem [{\citenamefont {Chi}\ \emph {et~al.}(2007)\citenamefont {Chi},
  \citenamefont {Choi}, \citenamefont {Kim}, \citenamefont {Kim},\ and\
  \citenamefont {Lee}}]{PhysRevA.75.032306}%
  \BibitemOpen
  \bibfield  {author} {\bibinfo {author} {\bibfnamefont {D.~P.}\ \bibnamefont
  {Chi}}, \bibinfo {author} {\bibfnamefont {J.~W.}\ \bibnamefont {Choi}},
  \bibinfo {author} {\bibfnamefont {J.~S.}\ \bibnamefont {Kim}}, \bibinfo
  {author} {\bibfnamefont {T.}~\bibnamefont {Kim}}, \ and\ \bibinfo {author}
  {\bibfnamefont {S.}~\bibnamefont {Lee}},\ }\href {\doibase
  10.1103/PhysRevA.75.032306} {\bibfield  {journal} {\bibinfo  {journal} {Phys.
  Rev. A}\ }\textbf {\bibinfo {volume} {75}},\ \bibinfo {pages} {032306}
  (\bibinfo {year} {2007})}\BibitemShut {NoStop}%
\bibitem [{\citenamefont {Mishra}\ \emph {et~al.}(2020)\citenamefont {Mishra},
  \citenamefont {Sengupta},\ and\ \citenamefont
  {Arvind}}]{PhysRevA.102.032415}%
  \BibitemOpen
  \bibfield  {author} {\bibinfo {author} {\bibfnamefont {M.}~\bibnamefont
  {Mishra}}, \bibinfo {author} {\bibfnamefont {R.}~\bibnamefont {Sengupta}}, \
  and\ \bibinfo {author} {\bibnamefont {Arvind}},\ }\href {\doibase
  10.1103/PhysRevA.102.032415} {\bibfield  {journal} {\bibinfo  {journal}
  {Phys. Rev. A}\ }\textbf {\bibinfo {volume} {102}},\ \bibinfo {pages}
  {032415} (\bibinfo {year} {2020})}\BibitemShut {NoStop}%
\bibitem [{\citenamefont {T\'oth}\ and\ \citenamefont
  {V\'ertesi}(2018)}]{PhysRevLett.120.020506}%
  \BibitemOpen
  \bibfield  {author} {\bibinfo {author} {\bibfnamefont {G.}~\bibnamefont
  {T\'oth}}\ and\ \bibinfo {author} {\bibfnamefont {T.}~\bibnamefont
  {V\'ertesi}},\ }\href {\doibase 10.1103/PhysRevLett.120.020506} {\bibfield
  {journal} {\bibinfo  {journal} {Phys. Rev. Lett.}\ }\textbf {\bibinfo
  {volume} {120}},\ \bibinfo {pages} {020506} (\bibinfo {year}
  {2018})}\BibitemShut {NoStop}%
\bibitem [{\citenamefont {Navascués}\ \emph {et~al.}(2005)\citenamefont
  {Navascués}, \citenamefont {Bae}, \citenamefont {Cirac}, \citenamefont
  {Lewestein}, \citenamefont {Sanpera},\ and\ \citenamefont
  {Acín}}]{Navascu_s_2005}%
  \BibitemOpen
  \bibfield  {author} {\bibinfo {author} {\bibfnamefont {M.}~\bibnamefont
  {Navascués}}, \bibinfo {author} {\bibfnamefont {J.}~\bibnamefont {Bae}},
  \bibinfo {author} {\bibfnamefont {J.~I.}\ \bibnamefont {Cirac}}, \bibinfo
  {author} {\bibfnamefont {M.}~\bibnamefont {Lewestein}}, \bibinfo {author}
  {\bibfnamefont {A.}~\bibnamefont {Sanpera}}, \ and\ \bibinfo {author}
  {\bibfnamefont {A.}~\bibnamefont {Acín}},\ }\href {\doibase
  10.1103/physrevlett.94.010502} {\bibfield  {journal} {\bibinfo  {journal}
  {Physical Review Letters}\ }\textbf {\bibinfo {volume} {94}} (\bibinfo {year}
  {2005}),\ 10.1103/physrevlett.94.010502}\BibitemShut {NoStop}%
\bibitem [{\citenamefont {Navascu\'es}\ and\ \citenamefont
  {Ac\'{\i}n}(2005)}]{PhysRevA.72.012303}%
  \BibitemOpen
  \bibfield  {author} {\bibinfo {author} {\bibfnamefont {M.}~\bibnamefont
  {Navascu\'es}}\ and\ \bibinfo {author} {\bibfnamefont {A.}~\bibnamefont
  {Ac\'{\i}n}},\ }\href {\doibase 10.1103/PhysRevA.72.012303} {\bibfield
  {journal} {\bibinfo  {journal} {Phys. Rev. A}\ }\textbf {\bibinfo {volume}
  {72}},\ \bibinfo {pages} {012303} (\bibinfo {year} {2005})}\BibitemShut
  {NoStop}%
\bibitem [{\citenamefont {{Xiang, Shao-Hua}}\ and\ \citenamefont {{Song,
  Ke-Hui}}(2013)}]{refId0}%
  \BibitemOpen
  \bibfield  {author} {\bibinfo {author} {\bibnamefont {{Xiang, Shao-Hua}}}\
  and\ \bibinfo {author} {\bibnamefont {{Song, Ke-Hui}}},\ }\href {\doibase
  10.1140/epjd/e2013-40009-6} {\bibfield  {journal} {\bibinfo  {journal} {Eur.
  Phys. J. D}\ }\textbf {\bibinfo {volume} {67}},\ \bibinfo {pages} {157}
  (\bibinfo {year} {2013})}\BibitemShut {NoStop}%
\bibitem [{\citenamefont {Marian}\ \emph {et~al.}(2015)\citenamefont {Marian},
  \citenamefont {Ghiu},\ and\ \citenamefont {Marian}}]{marian2015decay}%
  \BibitemOpen
  \bibfield  {author} {\bibinfo {author} {\bibfnamefont {P.}~\bibnamefont
  {Marian}}, \bibinfo {author} {\bibfnamefont {I.}~\bibnamefont {Ghiu}}, \ and\
  \bibinfo {author} {\bibfnamefont {T.~A.}\ \bibnamefont {Marian}},\
  }\href@noop {} {\bibfield  {journal} {\bibinfo  {journal} {Physica Scripta}\
  }\textbf {\bibinfo {volume} {90}},\ \bibinfo {pages} {074041} (\bibinfo
  {year} {2015})}\BibitemShut {NoStop}%
\bibitem [{\citenamefont {Xiang}\ \emph {et~al.}(2011)\citenamefont {Xiang},
  \citenamefont {Song}, \citenamefont {Wen},\ and\ \citenamefont
  {Shi}}]{xiang2011}%
  \BibitemOpen
  \bibfield  {author} {\bibinfo {author} {\bibfnamefont {S.~H.}\ \bibnamefont
  {Xiang}}, \bibinfo {author} {\bibfnamefont {K.~H.}\ \bibnamefont {Song}},
  \bibinfo {author} {\bibfnamefont {W.}~\bibnamefont {Wen}}, \ and\ \bibinfo
  {author} {\bibfnamefont {Z.~G.}\ \bibnamefont {Shi}},\ }\href@noop {}
  {\bibfield  {journal} {\bibinfo  {journal} {The European Physical Journal D}\
  }\textbf {\bibinfo {volume} {62}},\ \bibinfo {pages} {289} (\bibinfo {year}
  {2011})}\BibitemShut {NoStop}%
\bibitem [{\citenamefont {Xiang}\ \emph {et~al.}(2008)\citenamefont {Xiang},
  \citenamefont {Shao},\ and\ \citenamefont {Song}}]{PhysRevA.78.052313}%
  \BibitemOpen
  \bibfield  {author} {\bibinfo {author} {\bibfnamefont {S.-H.}\ \bibnamefont
  {Xiang}}, \bibinfo {author} {\bibfnamefont {B.}~\bibnamefont {Shao}}, \ and\
  \bibinfo {author} {\bibfnamefont {K.-H.}\ \bibnamefont {Song}},\ }\href
  {\doibase 10.1103/PhysRevA.78.052313} {\bibfield  {journal} {\bibinfo
  {journal} {Phys. Rev. A}\ }\textbf {\bibinfo {volume} {78}},\ \bibinfo
  {pages} {052313} (\bibinfo {year} {2008})}\BibitemShut {NoStop}%
\bibitem [{\citenamefont {Barbosa}\ \emph {et~al.}(2011)\citenamefont
  {Barbosa}, \citenamefont {De~Faria}, \citenamefont {Coelho}, \citenamefont
  {Cassemiro}, \citenamefont {Villar}, \citenamefont {Nussenzveig},\ and\
  \citenamefont {Martinelli}}]{barbosa2011disentanglement}%
  \BibitemOpen
  \bibfield  {author} {\bibinfo {author} {\bibfnamefont {F.}~\bibnamefont
  {Barbosa}}, \bibinfo {author} {\bibfnamefont {A.}~\bibnamefont {De~Faria}},
  \bibinfo {author} {\bibfnamefont {A.}~\bibnamefont {Coelho}}, \bibinfo
  {author} {\bibfnamefont {K.}~\bibnamefont {Cassemiro}}, \bibinfo {author}
  {\bibfnamefont {A.}~\bibnamefont {Villar}}, \bibinfo {author} {\bibfnamefont
  {P.}~\bibnamefont {Nussenzveig}}, \ and\ \bibinfo {author} {\bibfnamefont
  {M.}~\bibnamefont {Martinelli}},\ }\href@noop {} {\bibfield  {journal}
  {\bibinfo  {journal} {Physical Review A—Atomic, Molecular, and Optical
  Physics}\ }\textbf {\bibinfo {volume} {84}},\ \bibinfo {pages} {052330}
  (\bibinfo {year} {2011})}\BibitemShut {NoStop}%
\bibitem [{\citenamefont {Isar}(2011)}]{Isar2011}%
  \BibitemOpen
  \bibfield  {author} {\bibinfo {author} {\bibfnamefont {A.}~\bibnamefont
  {Isar}},\ }\href {\doibase 10.1088/0031-8949/2011/t143/014012} {\bibfield
  {journal} {\bibinfo  {journal} {Physica Scripta}\ }\textbf {\bibinfo {volume}
  {T143}},\ \bibinfo {pages} {014012} (\bibinfo {year} {2011})}\BibitemShut
  {NoStop}%
\bibitem [{\citenamefont {Rishabh}\ \emph {et~al.}(2022)\citenamefont
  {Rishabh}, \citenamefont {Kumar}, \citenamefont {Narang},\ and\ \citenamefont
  {Arvind}}]{PhysRevA.105.042405}%
  \BibitemOpen
  \bibfield  {author} {\bibinfo {author} {\bibnamefont {Rishabh}}, \bibinfo
  {author} {\bibfnamefont {C.}~\bibnamefont {Kumar}}, \bibinfo {author}
  {\bibfnamefont {G.}~\bibnamefont {Narang}}, \ and\ \bibinfo {author}
  {\bibnamefont {Arvind}},\ }\href {\doibase 10.1103/PhysRevA.105.042405}
  {\bibfield  {journal} {\bibinfo  {journal} {Phys. Rev. A}\ }\textbf {\bibinfo
  {volume} {105}},\ \bibinfo {pages} {042405} (\bibinfo {year}
  {2022})}\BibitemShut {NoStop}%
\bibitem [{\citenamefont {Tserkis}\ and\ \citenamefont
  {Ralph}(2017)}]{PhysRevA.96.062338}%
  \BibitemOpen
  \bibfield  {author} {\bibinfo {author} {\bibfnamefont {S.}~\bibnamefont
  {Tserkis}}\ and\ \bibinfo {author} {\bibfnamefont {T.~C.}\ \bibnamefont
  {Ralph}},\ }\href {\doibase 10.1103/PhysRevA.96.062338} {\bibfield  {journal}
  {\bibinfo  {journal} {Phys. Rev. A}\ }\textbf {\bibinfo {volume} {96}},\
  \bibinfo {pages} {062338} (\bibinfo {year} {2017})}\BibitemShut {NoStop}%
\bibitem [{\citenamefont {Roy}\ \emph {et~al.}(2020)\citenamefont {Roy},
  \citenamefont {Das},\ and\ \citenamefont {Sen(De)}}]{PhysRevA.102.012421}%
  \BibitemOpen
  \bibfield  {author} {\bibinfo {author} {\bibfnamefont {S.}~\bibnamefont
  {Roy}}, \bibinfo {author} {\bibfnamefont {T.}~\bibnamefont {Das}}, \ and\
  \bibinfo {author} {\bibfnamefont {A.}~\bibnamefont {Sen(De)}},\ }\href
  {\doibase 10.1103/PhysRevA.102.012421} {\bibfield  {journal} {\bibinfo
  {journal} {Phys. Rev. A}\ }\textbf {\bibinfo {volume} {102}},\ \bibinfo
  {pages} {012421} (\bibinfo {year} {2020})}\BibitemShut {NoStop}%
\bibitem [{\citenamefont {Hyllus}\ and\ \citenamefont
  {Eisert}(2006)}]{Hyllus_2006}%
  \BibitemOpen
  \bibfield  {author} {\bibinfo {author} {\bibfnamefont {P.}~\bibnamefont
  {Hyllus}}\ and\ \bibinfo {author} {\bibfnamefont {J.}~\bibnamefont
  {Eisert}},\ }\href {\doibase 10.1088/1367-2630/8/4/051} {\bibfield  {journal}
  {\bibinfo  {journal} {New Journal of Physics}\ }\textbf {\bibinfo {volume}
  {8}},\ \bibinfo {pages} {51–51} (\bibinfo {year} {2006})}\BibitemShut
  {NoStop}%
\bibitem [{\citenamefont {Mihaescu}\ \emph {et~al.}(2020)\citenamefont
  {Mihaescu}, \citenamefont {Kampermann}, \citenamefont {Gianfelici},
  \citenamefont {Isar},\ and\ \citenamefont {Bruß}}]{Mihaescu_2020}%
  \BibitemOpen
  \bibfield  {author} {\bibinfo {author} {\bibfnamefont {T.}~\bibnamefont
  {Mihaescu}}, \bibinfo {author} {\bibfnamefont {H.}~\bibnamefont
  {Kampermann}}, \bibinfo {author} {\bibfnamefont {G.}~\bibnamefont
  {Gianfelici}}, \bibinfo {author} {\bibfnamefont {A.}~\bibnamefont {Isar}}, \
  and\ \bibinfo {author} {\bibfnamefont {D.}~\bibnamefont {Bruß}},\ }\href
  {\doibase 10.1088/1367-2630/abd1ad} {\bibfield  {journal} {\bibinfo
  {journal} {New Journal of Physics}\ }\textbf {\bibinfo {volume} {22}},\
  \bibinfo {pages} {123041} (\bibinfo {year} {2020})}\BibitemShut {NoStop}%
\bibitem [{\citenamefont {Tavakoli}\ \emph {et~al.}(2024)\citenamefont
  {Tavakoli}, \citenamefont {Pozas-Kerstjens}, \citenamefont {Brown},\ and\
  \citenamefont {Araújo}}]{Tavakoli_2024}%
  \BibitemOpen
  \bibfield  {author} {\bibinfo {author} {\bibfnamefont {A.}~\bibnamefont
  {Tavakoli}}, \bibinfo {author} {\bibfnamefont {A.}~\bibnamefont
  {Pozas-Kerstjens}}, \bibinfo {author} {\bibfnamefont {P.}~\bibnamefont
  {Brown}}, \ and\ \bibinfo {author} {\bibfnamefont {M.}~\bibnamefont
  {Araújo}},\ }\href {\doibase 10.1103/revmodphys.96.045006} {\bibfield
  {journal} {\bibinfo  {journal} {Reviews of Modern Physics}\ }\textbf
  {\bibinfo {volume} {96}} (\bibinfo {year} {2024}),\
  10.1103/revmodphys.96.045006}\BibitemShut {NoStop}%
\bibitem [{\citenamefont {Ma}\ \emph {et~al.}(2020)\citenamefont {Ma},
  \citenamefont {Xue}, \citenamefont {Guo},\ and\ \citenamefont
  {Shu}}]{Ma_2020}%
  \BibitemOpen
  \bibfield  {author} {\bibinfo {author} {\bibfnamefont {S.}~\bibnamefont
  {Ma}}, \bibinfo {author} {\bibfnamefont {S.}~\bibnamefont {Xue}}, \bibinfo
  {author} {\bibfnamefont {Y.}~\bibnamefont {Guo}}, \ and\ \bibinfo {author}
  {\bibfnamefont {C.-C.}\ \bibnamefont {Shu}},\ }\href {\doibase
  10.1007/s11128-020-02726-1} {\bibfield  {journal} {\bibinfo  {journal}
  {Quantum Information Processing}\ }\textbf {\bibinfo {volume} {19}} (\bibinfo
  {year} {2020}),\ 10.1007/s11128-020-02726-1}\BibitemShut {NoStop}%
\bibitem [{\citenamefont {Rajarama~Bhat}\ \emph {et~al.}(2017)\citenamefont
  {Rajarama~Bhat}, \citenamefont {Parthasarathy},\ and\ \citenamefont
  {Sengupta}}]{Rajarama_Bhat_2017}%
  \BibitemOpen
  \bibfield  {author} {\bibinfo {author} {\bibfnamefont {B.~V.}\ \bibnamefont
  {Rajarama~Bhat}}, \bibinfo {author} {\bibfnamefont {K.~R.}\ \bibnamefont
  {Parthasarathy}}, \ and\ \bibinfo {author} {\bibfnamefont {R.}~\bibnamefont
  {Sengupta}},\ }\href {\doibase 10.1142/s0129055x1750012x} {\bibfield
  {journal} {\bibinfo  {journal} {Reviews in Mathematical Physics}\ }\textbf
  {\bibinfo {volume} {29}},\ \bibinfo {pages} {1750012} (\bibinfo {year}
  {2017})}\BibitemShut {NoStop}%
\bibitem [{\citenamefont {Leppäjärvi}\ \emph {et~al.}(2024)\citenamefont
  {Leppäjärvi}, \citenamefont {Nechita},\ and\ \citenamefont
  {Sengupta}}]{Lepp_j_rvi_2024}%
  \BibitemOpen
  \bibfield  {author} {\bibinfo {author} {\bibfnamefont {L.}~\bibnamefont
  {Leppäjärvi}}, \bibinfo {author} {\bibfnamefont {I.}~\bibnamefont
  {Nechita}}, \ and\ \bibinfo {author} {\bibfnamefont {R.}~\bibnamefont
  {Sengupta}},\ }\href {\doibase 10.1063/5.0202147} {\bibfield  {journal}
  {\bibinfo  {journal} {Journal of Mathematical Physics}\ }\textbf {\bibinfo
  {volume} {65}} (\bibinfo {year} {2024}),\ 10.1063/5.0202147}\BibitemShut
  {NoStop}%
\bibitem [{\citenamefont {Thomas}\ \emph {et~al.}(2017)\citenamefont {Thomas},
  \citenamefont {Bohmann},\ and\ \citenamefont {Vogel}}]{vogel}%
  \BibitemOpen
  \bibfield  {author} {\bibinfo {author} {\bibfnamefont {P.}~\bibnamefont
  {Thomas}}, \bibinfo {author} {\bibfnamefont {M.}~\bibnamefont {Bohmann}}, \
  and\ \bibinfo {author} {\bibfnamefont {W.}~\bibnamefont {Vogel}},\ }\href
  {\doibase 10.1103/PhysRevA.96.042321} {\bibfield  {journal} {\bibinfo
  {journal} {Phys. Rev. A}\ }\textbf {\bibinfo {volume} {96}},\ \bibinfo
  {pages} {042321} (\bibinfo {year} {2017})}\BibitemShut {NoStop}%
\bibitem [{\citenamefont {Williamson}(1936)}]{williamson1936algebraic}%
  \BibitemOpen
  \bibfield  {author} {\bibinfo {author} {\bibfnamefont {J.}~\bibnamefont
  {Williamson}},\ }\href@noop {} {\bibfield  {journal} {\bibinfo  {journal}
  {American journal of mathematics}\ }\textbf {\bibinfo {volume} {58}},\
  \bibinfo {pages} {141} (\bibinfo {year} {1936})}\BibitemShut {NoStop}%
\bibitem [{\citenamefont {Horodecki}(1997)}]{Horodecki1997separability}%
  \BibitemOpen
  \bibfield  {author} {\bibinfo {author} {\bibfnamefont {P.}~\bibnamefont
  {Horodecki}},\ }\href {\doibase 10.1016/s0375-9601(97)00416-7} {\bibfield
  {journal} {\bibinfo  {journal} {Physics Letters A}\ }\textbf {\bibinfo
  {volume} {232}},\ \bibinfo {pages} {333–339} (\bibinfo {year}
  {1997})}\BibitemShut {NoStop}%
\bibitem [{\citenamefont {Horodecki}\ \emph {et~al.}(1998)\citenamefont
  {Horodecki}, \citenamefont {Horodecki},\ and\ \citenamefont
  {Horodecki}}]{horodecki1998mixed}%
  \BibitemOpen
  \bibfield  {author} {\bibinfo {author} {\bibfnamefont {M.}~\bibnamefont
  {Horodecki}}, \bibinfo {author} {\bibfnamefont {P.}~\bibnamefont
  {Horodecki}}, \ and\ \bibinfo {author} {\bibfnamefont {R.}~\bibnamefont
  {Horodecki}},\ }\href {\doibase 10.1103/PhysRevLett.80.5239} {\bibfield
  {journal} {\bibinfo  {journal} {Phys. Rev. Lett.}\ }\textbf {\bibinfo
  {volume} {80}},\ \bibinfo {pages} {5239} (\bibinfo {year}
  {1998})}\BibitemShut {NoStop}%
\bibitem [{\citenamefont {Horodecki}\ \emph {et~al.}(1996)\citenamefont
  {Horodecki}, \citenamefont {Horodecki},\ and\ \citenamefont
  {Horodecki}}]{Horodecki1996separability}%
  \BibitemOpen
  \bibfield  {author} {\bibinfo {author} {\bibfnamefont {M.}~\bibnamefont
  {Horodecki}}, \bibinfo {author} {\bibfnamefont {P.}~\bibnamefont
  {Horodecki}}, \ and\ \bibinfo {author} {\bibfnamefont {R.}~\bibnamefont
  {Horodecki}},\ }\href {\doibase 10.1016/s0375-9601(96)00706-2} {\bibfield
  {journal} {\bibinfo  {journal} {Physics Letters A}\ }\textbf {\bibinfo
  {volume} {223}},\ \bibinfo {pages} {1–8} (\bibinfo {year}
  {1996})}\BibitemShut {NoStop}%
\bibitem [{\citenamefont {Giedke}\ \emph {et~al.}(2001)\citenamefont {Giedke},
  \citenamefont {Kraus}, \citenamefont {Lewenstein},\ and\ \citenamefont
  {Cirac}}]{Giedke_2001}%
  \BibitemOpen
  \bibfield  {author} {\bibinfo {author} {\bibfnamefont {G.}~\bibnamefont
  {Giedke}}, \bibinfo {author} {\bibfnamefont {B.}~\bibnamefont {Kraus}},
  \bibinfo {author} {\bibfnamefont {M.}~\bibnamefont {Lewenstein}}, \ and\
  \bibinfo {author} {\bibfnamefont {J.~I.}\ \bibnamefont {Cirac}},\ }\href
  {\doibase 10.1103/physrevlett.87.167904} {\bibfield  {journal} {\bibinfo
  {journal} {Physical Review Letters}\ }\textbf {\bibinfo {volume} {87}}
  (\bibinfo {year} {2001}),\ 10.1103/physrevlett.87.167904}\BibitemShut
  {NoStop}%
\bibitem [{\citenamefont {Serafini}\ \emph {et~al.}(2005)\citenamefont
  {Serafini}, \citenamefont {Adesso},\ and\ \citenamefont
  {Illuminati}}]{Serafini_2005}%
  \BibitemOpen
  \bibfield  {author} {\bibinfo {author} {\bibfnamefont {A.}~\bibnamefont
  {Serafini}}, \bibinfo {author} {\bibfnamefont {G.}~\bibnamefont {Adesso}}, \
  and\ \bibinfo {author} {\bibfnamefont {F.}~\bibnamefont {Illuminati}},\
  }\href {\doibase 10.1103/physreva.71.032349} {\bibfield  {journal} {\bibinfo
  {journal} {Physical Review A}\ }\textbf {\bibinfo {volume} {71}} (\bibinfo
  {year} {2005}),\ 10.1103/physreva.71.032349}\BibitemShut {NoStop}%
\bibitem [{\citenamefont {Lami}\ \emph {et~al.}(2019)\citenamefont {Lami},
  \citenamefont {Khatri}, \citenamefont {Adesso},\ and\ \citenamefont
  {Wilde}}]{PhysRevLett.123.050501}%
  \BibitemOpen
  \bibfield  {author} {\bibinfo {author} {\bibfnamefont {L.}~\bibnamefont
  {Lami}}, \bibinfo {author} {\bibfnamefont {S.}~\bibnamefont {Khatri}},
  \bibinfo {author} {\bibfnamefont {G.}~\bibnamefont {Adesso}}, \ and\ \bibinfo
  {author} {\bibfnamefont {M.~M.}\ \bibnamefont {Wilde}},\ }\href {\doibase
  10.1103/PhysRevLett.123.050501} {\bibfield  {journal} {\bibinfo  {journal}
  {Phys. Rev. Lett.}\ }\textbf {\bibinfo {volume} {123}},\ \bibinfo {pages}
  {050501} (\bibinfo {year} {2019})}\BibitemShut {NoStop}%
\bibitem [{\citenamefont {Gardiner}\ and\ \citenamefont
  {Collett}(1985)}]{gardiner}%
  \BibitemOpen
  \bibfield  {author} {\bibinfo {author} {\bibfnamefont {C.~W.}\ \bibnamefont
  {Gardiner}}\ and\ \bibinfo {author} {\bibfnamefont {M.~J.}\ \bibnamefont
  {Collett}},\ }\href {\doibase 10.1103/PhysRevA.31.3761} {\bibfield  {journal}
  {\bibinfo  {journal} {Phys. Rev. A}\ }\textbf {\bibinfo {volume} {31}},\
  \bibinfo {pages} {3761} (\bibinfo {year} {1985})}\BibitemShut {NoStop}%
\bibitem [{\citenamefont {Olivares}(2012)}]{Olivares_2012}%
  \BibitemOpen
  \bibfield  {author} {\bibinfo {author} {\bibfnamefont {S.}~\bibnamefont
  {Olivares}},\ }\href {\doibase 10.1140/epjst/e2012-01532-4} {\bibfield
  {journal} {\bibinfo  {journal} {The European Physical Journal Special
  Topics}\ }\textbf {\bibinfo {volume} {203}},\ \bibinfo {pages} {3–24}
  (\bibinfo {year} {2012})}\BibitemShut {NoStop}%
\bibitem [{\citenamefont {Ma}\ and\ \citenamefont {Rhodes}(1990)}]{Ma1990}%
  \BibitemOpen
  \bibfield  {author} {\bibinfo {author} {\bibfnamefont {X.}~\bibnamefont
  {Ma}}\ and\ \bibinfo {author} {\bibfnamefont {W.}~\bibnamefont {Rhodes}},\
  }\href {\doibase 10.1103/PhysRevA.41.4625} {\bibfield  {journal} {\bibinfo
  {journal} {Phys. Rev. A}\ }\textbf {\bibinfo {volume} {41}},\ \bibinfo
  {pages} {4625} (\bibinfo {year} {1990})}\BibitemShut {NoStop}%
\bibitem [{\citenamefont {Das}\ \emph {et~al.}(2016)\citenamefont {Das},
  \citenamefont {Prabhu}, \citenamefont {Sen(De)},\ and\ \citenamefont
  {Sen}}]{Das2016}%
  \BibitemOpen
  \bibfield  {author} {\bibinfo {author} {\bibfnamefont {T.}~\bibnamefont
  {Das}}, \bibinfo {author} {\bibfnamefont {R.}~\bibnamefont {Prabhu}},
  \bibinfo {author} {\bibfnamefont {A.}~\bibnamefont {Sen(De)}}, \ and\
  \bibinfo {author} {\bibfnamefont {U.}~\bibnamefont {Sen}},\ }\href {\doibase
  10.1103/PhysRevA.93.052313} {\bibfield  {journal} {\bibinfo  {journal} {Phys.
  Rev. A}\ }\textbf {\bibinfo {volume} {93}},\ \bibinfo {pages} {052313}
  (\bibinfo {year} {2016})}\BibitemShut {NoStop}%
\bibitem [{\citenamefont {Ma}\ \emph {et~al.}(2024)\citenamefont {Ma},
  \citenamefont {Zhou}, \citenamefont {Ma},\ and\ \citenamefont
  {Xue}}]{10661902}%
  \BibitemOpen
  \bibfield  {author} {\bibinfo {author} {\bibfnamefont {S.}~\bibnamefont
  {Ma}}, \bibinfo {author} {\bibfnamefont {L.}~\bibnamefont {Zhou}}, \bibinfo
  {author} {\bibfnamefont {J.}~\bibnamefont {Ma}}, \ and\ \bibinfo {author}
  {\bibfnamefont {S.}~\bibnamefont {Xue}},\ }in\ \href {\doibase
  10.23919/CCC63176.2024.10661902} {\emph {\bibinfo {booktitle} {2024 43rd
  Chinese Control Conference (CCC)}}}\ (\bibinfo {year} {2024})\ pp.\ \bibinfo
  {pages} {6777--6782}\BibitemShut {NoStop}%
\bibitem [{\citenamefont {Adesso}\ \emph {et~al.}(2007)\citenamefont {Adesso},
  \citenamefont {Ericsson},\ and\ \citenamefont {Illuminati}}]{Adesso_2007}%
  \BibitemOpen
  \bibfield  {author} {\bibinfo {author} {\bibfnamefont {G.}~\bibnamefont
  {Adesso}}, \bibinfo {author} {\bibfnamefont {M.}~\bibnamefont {Ericsson}}, \
  and\ \bibinfo {author} {\bibfnamefont {F.}~\bibnamefont {Illuminati}},\
  }\href {\doibase 10.1103/physreva.76.022315} {\bibfield  {journal} {\bibinfo
  {journal} {Physical Review A}\ }\textbf {\bibinfo {volume} {76}} (\bibinfo
  {year} {2007}),\ 10.1103/physreva.76.022315}\BibitemShut {NoStop}%
\bibitem [{\citenamefont {Simon}\ \emph {et~al.}(1994)\citenamefont {Simon},
  \citenamefont {Mukunda},\ and\ \citenamefont {Dutta}}]{Simon1994}%
  \BibitemOpen
  \bibfield  {author} {\bibinfo {author} {\bibfnamefont {R.}~\bibnamefont
  {Simon}}, \bibinfo {author} {\bibfnamefont {N.}~\bibnamefont {Mukunda}}, \
  and\ \bibinfo {author} {\bibfnamefont {B.}~\bibnamefont {Dutta}},\ }\href
  {\doibase 10.1103/PhysRevA.49.1567} {\bibfield  {journal} {\bibinfo
  {journal} {Phys. Rev. A}\ }\textbf {\bibinfo {volume} {49}},\ \bibinfo
  {pages} {1567} (\bibinfo {year} {1994})}\BibitemShut {NoStop}%
\bibitem [{\citenamefont {Roy}(2024)}]{roytypical2024}%
  \BibitemOpen
  \bibfield  {author} {\bibinfo {author} {\bibfnamefont {S.}~\bibnamefont
  {Roy}},\ }\href {\doibase 10.48550/ARXIV.2404.17265} {\enquote {\bibinfo
  {title} {Typical behaviour of genuine multimode entanglement of pure gaussian
  states},}\ } (\bibinfo {year} {2024})\BibitemShut {NoStop}%
\bibitem [{\citenamefont {Ma}\ \emph {et~al.}(2019)\citenamefont {Ma},
  \citenamefont {Woolley}, \citenamefont {Jia},\ and\ \citenamefont
  {Zhang}}]{Ma_2019}%
  \BibitemOpen
  \bibfield  {author} {\bibinfo {author} {\bibfnamefont {S.}~\bibnamefont
  {Ma}}, \bibinfo {author} {\bibfnamefont {M.~J.}\ \bibnamefont {Woolley}},
  \bibinfo {author} {\bibfnamefont {X.}~\bibnamefont {Jia}}, \ and\ \bibinfo
  {author} {\bibfnamefont {J.}~\bibnamefont {Zhang}},\ }\href {\doibase
  10.1103/physreva.100.022309} {\bibfield  {journal} {\bibinfo  {journal}
  {Physical Review A}\ }\textbf {\bibinfo {volume} {100}} (\bibinfo {year}
  {2019}),\ 10.1103/physreva.100.022309}\BibitemShut {NoStop}%
\bibitem [{\citenamefont {Chen}\ \emph {et~al.}(2023)\citenamefont {Chen},
  \citenamefont {Miao}, \citenamefont {Yin},\ and\ \citenamefont
  {Yuan}}]{PhysRevA.107.022410}%
  \BibitemOpen
  \bibfield  {author} {\bibinfo {author} {\bibfnamefont {X.-y.}\ \bibnamefont
  {Chen}}, \bibinfo {author} {\bibfnamefont {M.}~\bibnamefont {Miao}}, \bibinfo
  {author} {\bibfnamefont {R.}~\bibnamefont {Yin}}, \ and\ \bibinfo {author}
  {\bibfnamefont {J.}~\bibnamefont {Yuan}},\ }\href {\doibase
  10.1103/PhysRevA.107.022410} {\bibfield  {journal} {\bibinfo  {journal}
  {Phys. Rev. A}\ }\textbf {\bibinfo {volume} {107}},\ \bibinfo {pages}
  {022410} (\bibinfo {year} {2023})}\BibitemShut {NoStop}%
\bibitem [{\citenamefont {An}\ and\ \citenamefont {Zhang}(2007)}]{An2007}%
  \BibitemOpen
  \bibfield  {author} {\bibinfo {author} {\bibfnamefont {J.-H.}\ \bibnamefont
  {An}}\ and\ \bibinfo {author} {\bibfnamefont {W.-M.}\ \bibnamefont {Zhang}},\
  }\href {\doibase 10.1103/PhysRevA.76.042127} {\bibfield  {journal} {\bibinfo
  {journal} {Phys. Rev. A}\ }\textbf {\bibinfo {volume} {76}},\ \bibinfo
  {pages} {042127} (\bibinfo {year} {2007})}\BibitemShut {NoStop}%
\end{thebibliography}
%

\end{document}